\documentclass[twocolumn]{pasj00}

\begin{document}
\SetRunningHead{Tanaka et al.}{H$\alpha$ Emitters around 4C~23.56}

\title{Discovery of an Excess of H$\alpha$ Emitters around 4C~23.56 at ${\rm z}=2.48$}


\author{Ichi \textsc{Tanaka}\thanks{ichi@subaru.naoj.org}, \altaffilmark{1} 
Carlos \textsc{De Breuck}, \altaffilmark{2}
Jaron D. \textsc{Kurk}, \altaffilmark{3}
Yoshiaki \textsc{Taniguchi}, \altaffilmark{4}}
\author{Tadayuki \textsc{Kodama}, \altaffilmark{1,5}
Yuichi \textsc{Matsuda}, \altaffilmark{6}
Chris \textsc{Packham}, \altaffilmark{7}
Andrew \textsc{Zirm}, \altaffilmark{8}
Masaru \textsc{Kajisawa}, \altaffilmark{5,9}}
\author{Takashi \textsc{Ichikawa}, \altaffilmark{9}
Nick \textsc{Seymour}, \altaffilmark{10}
Daniel \textsc{Stern}, \altaffilmark{11}
Alan \textsc{Stockton}, \altaffilmark{12}
Bram P. \textsc{Venemans}, \altaffilmark{2}
\and
Jo\"{e}l \textsc{Vernet} \altaffilmark{2}}


\altaffiltext{1}{Subaru Telescope, National Astronomical Observatory of Japan, 650 North A'ohoku Place, Hilo, HI 96720, USA}

\altaffiltext{2}{European Southern Observatory, Karl-Schwarzschild Strasse 2, 85748 Garching bei M{\"u}nchen, Germany}
\altaffiltext{3}{Max-Planck-Institut f\"{u}r Extraterrestrische Physik, Postfach 1312, 85741 Garching, Germany}
\altaffiltext{4}{Research Center for Space and Cosmic Evolution, Ehime University, Bunkyo-cho 2-5, Matsuyama 790-8577, Japan}
\altaffiltext{5}{National Astronomical Observatory of Japan, 2-21-1 Osawa, Mitaka, Tokyo 181-8588, Japan}
\altaffiltext{6}{Department of Physics, University of Durham, South Road, Durham DH1 3LE, UK}
\altaffiltext{7}{Department of Astronomy, University of Florida, 211 Bryant Science Center, P.O. Box 112055, Gainesville, FL 32611-2055, USA}
\altaffiltext{8}{Dark Cosmology Centre, Niels Bohr Institute, University of Copenhagen, Juliane Maries Vej 30, DK-2100 Copenhagen, Denmark}
\altaffiltext{9}{Astronomical Institute, Graduate School of Science, Tohoku University, Aramaki, Aoba, Sendai 980-8578, Japan}
\altaffiltext{10}{Mullard Space Science Laboratory, UCL, Holmbury St Mary, Dorking, Surrey, RH5 6NT, UK}
\altaffiltext{11}{Jet Propulsion Laboratory, California Institute of Technology, 4800 Oak Grove Dr., Pasadena, CA 91109, USA}
\altaffiltext{12}{Institute for Astronomy, University of Hawaii, Honolulu, HI 96822, USA}

\KeyWords{Galaxies: protocluster --- Galaxies: individual (4C~23.56) --- Galaxies: high redshifts --- Galaxies: H$\alpha$ emitters}

\maketitle

\begin{abstract}
  We report the discovery of a significant excess of candidate
  H$\alpha$ emitters (HAEs) in the field of the radio galaxy 4C~23.56 at
  ${\rm z}=2.483$. Using the MOIRCS near-infrared imager on the Subaru
  Telescope we found 11 candidate emission-line galaxies to a flux
  limit of $\sim7.5 \times 10^{-17} {\rm erg}\ {\rm s}^{-1}\ {\rm
    cm}^{-2}$, which is about 5 times excess from the expected field counts with 
  $\sim3$-$\sigma$ significance. Three of these are spectroscopically confirmed as
  redshifted H$\alpha$ at ${\rm z}=2.49$. The distribution of candidate emitters
  on the sky is tightly confined to a 1.2-Mpc-radius area at
  ${\rm z}=2.49$, locating 4C 23.56 at the western edge of the distribution. 
  Analysis of the deep {\it Spitzer} MIPS $24~\mu{\rm m}$ imaging
  shows that there is also an excess of faint MIPS sources. All but
  two of the 11 HAEs are also found in the MIPS data. 
  The inferred star-formation rate (SFR) of the HAEs based on the extinction-corrected 
H$\alpha$ luminosity (median SFR $\gtrsim100~{\rm M}_{\solar}~{\rm yr}^{-1}$) is 
similar to those of HAEs in random fields at ${\rm z}\sim2$. On the other hand, 
the MIPS-based SFR for the HAEs is on average 3.6 times larger, suggesting the 
existence of the star-formation significanly obscured by dust.
 The comparison of the H$\alpha$-based star-formation activities of
  the HAEs in the 4C~23.56 field to those in another proto-cluster around 
  PKS~1138-262 at ${\rm z}=2.16$ reveals that the latter tend to
  have fainter H$\alpha$ emission despite similar $K$-band magnitudes.
  This suggests that star-formation may be suppressed in the
  PKS~1138-262 protocluster relative to the 4C~23.56 protocluster.  This difference
  among the HAEs in the two proto-clusters at ${\rm z}>2$ may imply that
  some massive cluster galaxies are just forming at these epochs 
  with some variation among clusters.
\end{abstract}

\section{Introduction}
Clusters of galaxies are the most massive bound systems in the
universe. The present-day cluster population is dominated by
early-type galaxies which make up a tight ``red sequence'' in
color-magnitude diagram \citep{dressler80}. The analyses of the
zero-point, slope, and scatter of the color-magnitude relation have
been playing key roles in understanding when and how galaxies in
clusters are formed and evolved (e.g., \cite{bower92}).

Studies of low to intermediate redshift clusters indicate that the
majority of stars in cluster galaxies are formed at ${\rm z}>2$\ (e.g., \cite{ellis97,bower98,kodama98,stanford98,holden04,tran07}). On
the other hand, detailed morphological analyses of the cluster red
galaxies suggest that a significant fraction of the present-day
cluster early-type galaxies must have been fully assembled after
${\rm z}\sim1$ (e.g., \cite{dressler97, vandokkum00, vandokkum01, ellis06,
  mei09}). A significant rise in star-formation activity in the cluster
cores is also seen as one goes to higher ($z>0.4$) redshift, the
``Butcher-Oemler effect'' \citep{bo78,bo84}.

Studies of the Butcher-Oemler effect, as well as the morphological
transformation in cluster cores, have been among the major topics in
extragalactic astronomy for the past two decades (e.g., \cite{dressler94,couch94,dressler97,balogh97,oemler97,poggianti99,vandokkum00, ellingson01, kodama01,mei09, pannella09}). As the observational data
have been accumulated for ever more distant galaxy clusters, the ``break''
of the red sequence at fainter magnitudes and its evolution has
attracted great attention (e.g., \cite{kajisawa00,kodama04,delucia04,tanaka05,delucia07,gilbank08}). These
phenomena have been discussed in the framework of the ``down-sizing''
of the star-formation activity, whereby the stellar mass of the
galaxies which host star formation at a fixed rate is, on average, decreasing with
cosmic time. The evolution of the faint end of the red sequence is
also connected to the continuous infall of star-forming galaxies into
the cluster cores, their subsequent cessation of star-formation, and
change in their morphologies \citep{poggianti04, bundy06}.

From the local universe to intermediate redshifts (${\rm z}<1$), a
number of studies suggest that star formation is strongly suppressed
in the cluster cores compared with the low-density field environment (e.g., 
\cite{hashimoto98, gomez03, nakata05, poggianti06, patel09}).
For passively-evolving early-type galaxies, there is also weak, but
growing, evidence that the luminosity-weighted ages of galaxies tend
to be older in higher density environments. This suggests that perhaps
the time scale of star formation and/or secondary episodes of the star
formation vary with environment
\citep{thomas05, clemens06,rakos08,clemens09,blanton09,pannella09, thomas10}.
Recently, however, this trend is observed to flatten or even reverse,
with the density dependence of star-formation disappearing or with
enhanced star-formation in the cluster core at z$>1$ \citep{elbaz07,ideue09,hayashi10, tran10}.

If the majority of the massive cluster galaxies ($M_{\star} >
10^{10}~{\rm M}_{\solar}$) begin forming their stars at ${\rm z}>2$, they will have
high star formation rates at higher redshifts.  \citet{hayashi10} have
surveyed star-forming galaxies in the X-ray cluster XMMXCS J2215.9-1738
at ${\rm z}=1.46$
using a narrowband filter that samples the redshifted [O\emissiontype{II}] doublet
located at the cluster redshift. They showed that the [O\emissiontype{II}] emitting
star-forming galaxies are prevalent everywhere in the cluster
including its core region (see also, \cite{hilton10}).  Although star formation is 
still confined to the cluster outskirts in other $\rm{z}\sim1.4$ cluster 
\citep{lidman08, bauer10, strazzullo10}, the widespread star formation in XMMXCS J2215.9-1738 is in strong contrast to similar surveys of lower redshift clusters at ${\rm z}=0.81$ and ${\rm z}=0.41$ \citep{kodama04, koyama10}. 
The color-magnitude relation for XMMXCS J2215.9-1738 also shows a larger scatter than those of other intermediate redshift clusters \citep{hilton09,papovich10}. 
Although we need a larger sample of clusters at intermediate and high redshifts to 
discern a general trend in the star formation $-$ galaxy density relation \citep{elbaz07, poggianti08, sobral10}, the existence of widespread star formation activity in this $\rm{z}=1.5$ cluster strongly suggests that we are gradually approaching the major epoch of galaxy formation in cluster environments.

At higher redshifts (${\rm z}>2$), the number of the known (proto-) clusters
is still quite small, and the majority of the known proto-clusters
were discovered in narrowband Ly$\alpha$-emitter surveys.  With some
exceptions (e.g., \cite{steidel98,steidel00,shimasaku03,steidel05,gobat10}),
much of the search for proto-clusters has been done around powerful
radio galaxies and QSOs with known redshifts (e.g., \cite{keel99,
   pentericci00, kurk00, yamada01,venemans02, kurk04a, venemans05,
  venemans07, matsuda09}). The success of these searches around radio
galaxies is related to the fact that the host galaxies of the powerful
radio sources tend to be the most massive galaxies at any epoch of the
universe \citep{debreuck02,seymour2007}. Powerful radio sources are
powered by central super-massive black holes, and the large inferred
masses of the black holes in turn imply massive host stellar systems
\citep{magorrian98}. The largest galaxies in the universe tend to lie
in the centers of largest-scale haloes which must have originated from
the primordial density peaks where galaxy formation processes are
accelerated in a biased way \citep{rees85}.

Recently \citet{tanaka10} analyzed the broadband SED of galaxies in
the PKS~1138 proto-cluster and found that proto-cluster galaxies show
suppressed star-formation on average compared to field
galaxies (see also, \cite{zirm08, doherty10}). This implies that even
at ${\rm z}\gtrsim2$ the activity in the cluster core region seems to lag
behind the star-formation activity we see in galaxies in field
surveys. \citet{steidel00} also reported older stellar ages for their
${\rm z}=2.3$ proto-cluster than for field galaxies in the same survey,
implying a similar trend.  The detection of a possible
(proto-) red-sequence in four known proto-cluster fields at ${\rm z}\gtrsim2$
via $JHK$ photometry \citep{kodama07} also indicates that, though the
star-formation is also likely to be prevalent there, some $K$-bright
galaxies may already be converging to the red-sequence.

At higher redshifts, there are some hints that more intense
star-formation is occurring even in the proto-cluster environment. In
the SSA22 field, which is known to contain a large overdensity of
galaxies at ${\rm z}=3.1$ \citep{steidel98,steidel00}, \citet{matsuda04}
detected a significant number of the extended Ly$\alpha$ blobs which are
thought to be progenitors of massive galaxies (e.g., \cite{dijkstra09}). 
Two of the largest Ly$\alpha$ blobs are identified with red, massive 
($M_{\star} \sim 10^{11}~{\rm M}_{\solar}$) galaxies,
comparable to dominant cluster galaxies at ${\rm z}<1.5$\ \citep{uchimoto08,
 stott10}. \citet{tamura09} have detected an excess of AzTEC
$1.1$ mm-sources in the same SSA22 field.  The positional correlation
of the AzTEC sources with the Ly$\alpha$ emitter distribution at
${\rm z}=3.1$ suggests that dust-enshrouded vigorous star-formation might be
going on in the field. There is also a report of a significant excess
of submm-bright objects in the TNJ~1338-1942 field at ${\rm z}=4.1$, which
is known to have a proto-cluster identified by Ly$\alpha$ emitters
\citep{debreuck04, venemans07}.

As was demonstrated by \citet{kodama01}, star formation activity in
cluster cores should have been as vigorous as in the field environment
at some point in the past (and may exceed it at even higher redshifts).
To observe this crossover point we need larger samples of proto-clusters at
${\rm z}=$2--4.

The field around 4C~23.56 is one of the most promising
proto-cluster candidates at ${\rm z}>2$ (\cite{KC97}, hereafter KC97). The
authors detected an excess of candidate Ly$\alpha$ emitters as well as
an excess of very red objects in $I-K'$ color in their survey field
(1.25~arcmin$^{2}$).  More recently, \citet{kajisawa06a} took deep $J$
and $K_{\rm s}$ images with CISCO on Subaru and found a significant
excess of Distant Red Galaxies (DRGs, \cite{franx03}) in a
$1.6\times1.6$ arcmin$^{2}$ field.  This further supports the idea
that there are a number of evolved galaxies in this proto-cluster at
${\rm z}\sim2.49$.  However, both of these surveys have very limited field
coverage and they may not be probing the whole picture of the
proto-cluster.  So far no other follow-up observations, including any
redshift confirmation, have been reported in the literature.

Here we present the result of our new H$\alpha$ emitter (HAE) search
around 4C~23.56 with the wide-field near-infrared imager, MOIRCS,
mounted on the 8.2m Subaru Telescope. In this study, we show that
there is a clear and strong excess of the HAEs near the radio galaxy.
We present their star formation properties and discuss the
characteristics of this proto-cluster by comparing with other fields
at similar redshifts.  Archival {\it Spitzer} MIPS data are also
used to probe dusty star forming activity in the proto-cluster.  This
is the first narrow-band H$\alpha$ emitter survey of a proto-cluster
field at ${\rm z}>2$ after \citet{kurk04a}. The field size we cover is
nearly 19 times larger than that of KC97, and is more than twice as
large as that of \citet{kurk04a}.  We note that the survey redshift of
${\rm z}\sim2.5$ is close to the red end of the H$\alpha$-observable
window (${\rm z}<2.7$) from the ground, due to the significant rise of the
thermal background beyond the {\it K}-band window.

Throughout the paper, the standard
$\Lambda$CDM cosmology with $\Omega_{m}=0.3, \Omega_{\Lambda}=0.7,
h=0.7$ is assumed ($1\arcmin = 1.18$ comoving Mpc at ${\rm z}=2.48$). All
the magnitudes presented in this paper are in the AB system unless
otherwise noted.

\section{Observations and Data Reduction}

\subsection{Observations}
The primary data for the current work were collected using the
Multi-Object Infrared Camera and
Spectrograph (MOIRCS, \cite{ichikawa06, suzuki08}) mounted on the
Cassegrain focus of the Subaru 8.2-m Telescope at Mauna Kea. The
observations were executed during the second half-night of UT 1st June,
2007, during the author's (IT) Subaru Staff Time. The weather was
photometric but with variable seeing in the range
$\timeform{0.47''}-\timeform{0.7''}$.

MOIRCS has two independent sets of optics (including both filters and
detectors), which divide the full $4\arcmin\times7\arcmin$ field of
view (FOV) into the two subfields (channels 1 and 2), each covering
$4\arcmin\times\timeform{3.5'}$. We set the MOIRCS position angle to
$-90\arcdeg$ so that the long side ($=7\arcmin$ side) of the FOV aligns
to the east-west direction. In this setting, the channel 1 FOV is on
the eastern (left) side (see Fig.\ref{fig1}b).

We used both the standard $K_{\rm s}$-band filter and a narrowband
(NB) filter called ``CO'' or ``NB2288''. 
The NB filter has central wavelength $\lambda_{c}=2.2881~\mu$m with a
bandwidth of 0.023~$\mu$m.  Note that these are the post-installation
measured values (M. Akiyama, private communication). Individual NB
exposures were 300~s. In total, we obtained 16 exposures, yielding a
total integration time of 4800~s. Shallow data were also
acquired in the $K_{\rm s}$ band to enable continuum subtraction of
the NB data. 
The $K_{\rm s}$ band exposures were taken in 55~s and 3-coadds. We 
obtained 9 frames in total (1485~s). We found that stars on three
images were elongated due to poor guiding at high elevation ($>84$
degrees). We rejected those frames during the final combine, resulting
in a total integration of 990~s.  We applied a standard
circular-dither pattern with a diameter of $20\arcsec$.  After the
target observation, four position dithered data of the FS151 standard
star were acquired in both the $K_{\rm s}$ and NB2288 bands.

We note that 4C~23.56 is close ($\sim\timeform{1D}$) to a bright ($K\sim0$
mag) Mira-type star. Such bright stars are known to occasionally cause
significant stray light on the image\footnote{See the {\it Imaging
    Information} page on MOIRCS website.}. Fortunately, the
effect was small during the night.  However, stray light was seen in
some of the data. It affected only a tiny portion ($<1\%$) of the total
area, and we took care to mark the positions of any stray light before
identifying the emitters. 
Unfortunately, stray light caused a hole-like 
pattern near the radio galaxy in the $K_{\rm s}$ data.
Due to this, the total magnitude for
4C~23.56 may be systematically wrong. For other targets, we confirmed
that stray light does not affect the photometry.

\vspace{1ex}

\subsubsection{Wavelength Shift of NB filters}
In general, the central wavelength of NB filters will shift
with the angle of incidence of the light to the filter surface.  In
our case, therefore, the observable wavelength window will change with
the position of targets on the detector.
%
%

\begin{figure*}
\begin{center}
\FigureFile(90mm,100mm){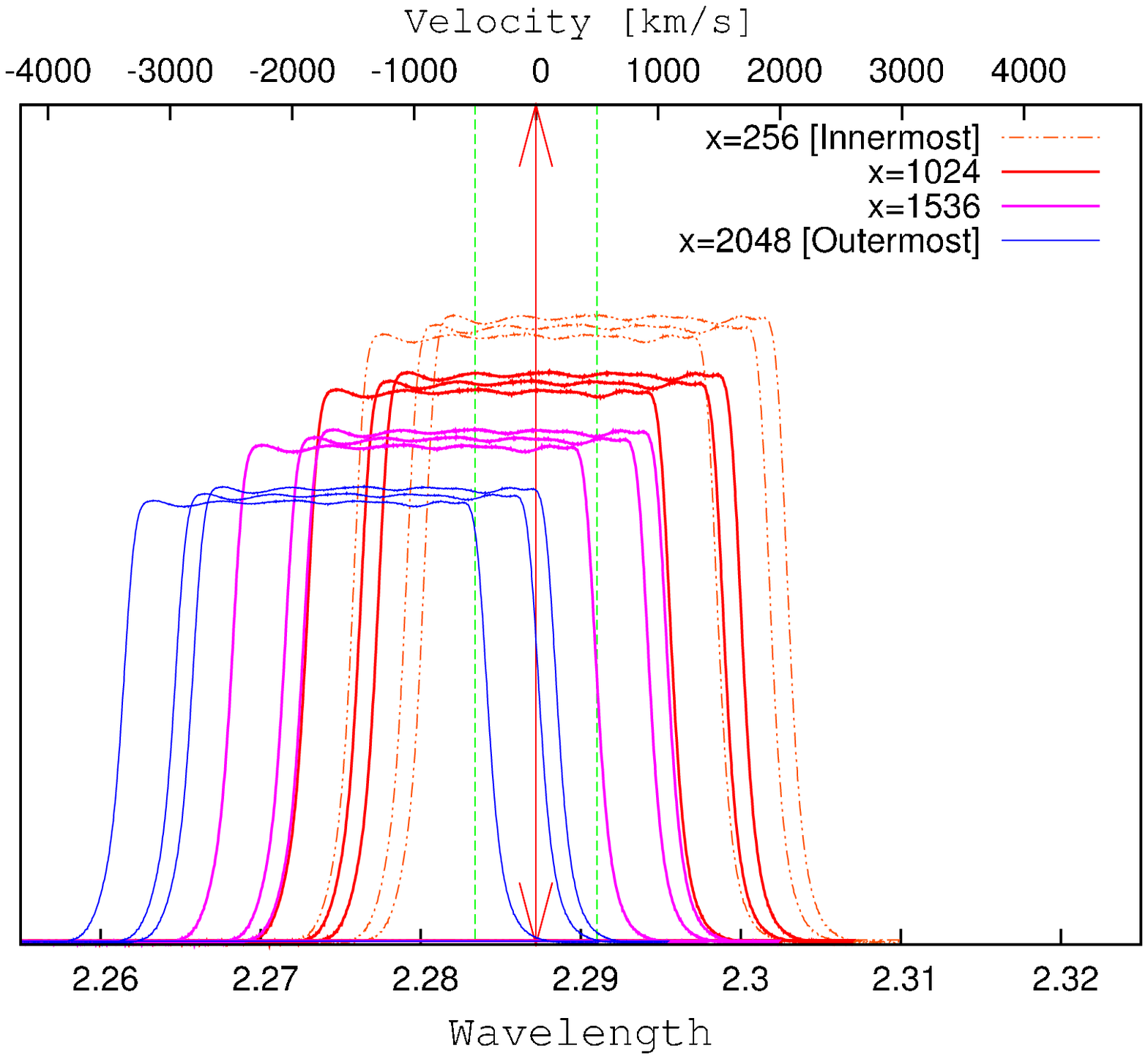}
\vspace{3ex}
\FigureFile(100mm,100mm){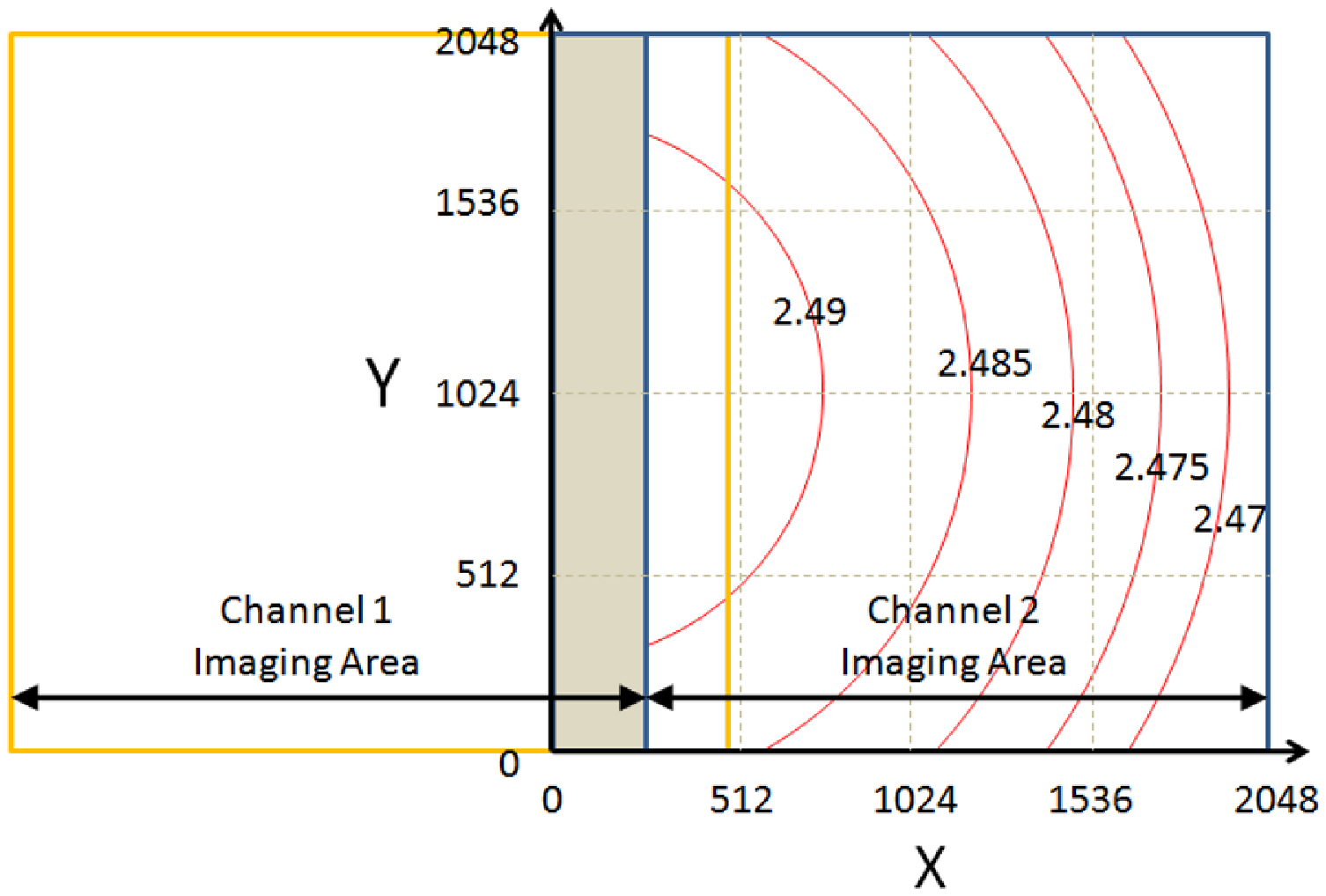}
\vspace{-0ex}
\end{center}
\caption{{\bf (Top)} The simulated shift of the transmission curve of the NB2289 filter. The horizontal axis is the wavelengths in $\mu$m and the vertical axis has arbitrary units. The curves with orange (dot-dot-dashed), red (thick solid), magenta (thick solid), and blue (thin solid) colors each indicate the transmission curves at $x$=256 (close to the innermost FOV coverage), 1024 (detector center), 1536, and 2048 (outer edge of the FOV) in channel-2 raw-image coordinates (the bottom figure, and corresponds to the right-half of the MOIRCS FOV). At each $x$ position, the shift of the transmission curve along with $y$-coordinates is also shown: the curve at the longest wavelength is at the detector center ($y$=1024), and shifts to shorter wavelengths as the $y$ coordinates goes to the edge of the detector ($y$=512/1536:middle and 1/2048:bluest). The red arrow indicates the expected position of the H$\alpha$ emission, and the vertical green lines (dots) indicate the range of the shift for objects at $\pm500~{\rm km}~{\rm s}^{-1}$ relative to the radio galaxy. The peak amplitude of the transmission curves at each $x$ coordinate are artificially shifted for clarity. See discussion for \S2.1.1 for details. {\bf (Bottom)} The schematic figure of the MOIRCS field of view for explanation of the top figure. The full $\timeform{4'}\times\timeform{7'}$ FOV is divided into two channels. Note that the minimum $x$ coordinate of the raw data area is around $x=270$ for channel~2 ($x<270$ is blocked by the mirror shadow). The curves on the channel-2 side show the iso-redshift contours (each number indicates the redshift) we can detect H$\alpha$ at the central wavelength of the filter window. The NB2288 filter can cover the redshift range of $\pm0.017$. The pointing center is around $(270,1024)$.\label{fig1}}
\end{figure*}

In Figure~\ref{fig1} we show the simulated shift of the filter
transmission curve with source position on the detector (the case for
channel 2 is shown here: R. Suzuki, private comm.). We find that the
redshift range of interest, $2.47 < {\rm z} < 2.49$, is always covered by
$>80\%$ of the detector area. As a reference, the expected wavelength
of the H$\alpha$ emission from 4C~23.56 is plotted as the arrow in
Fig.\ref{fig1}. Note that the redshift of 4C~23.56 is $2.483\pm0.003$
by \citet{rot97} from C\emissiontype{IV}, He\emissiontype{II}, and C\emissiontype{III}]
lines. The bottom of Fig.\ref{fig1} shows the iso-redshift lines of
the observable H$\alpha$ emission for the center of the filter
transmission. The iso-redshift line for 2.483 goes across the center
of the detector.  In the upper panel of Fig.\ref{fig1} we show that
H$\alpha$ emission at $\pm500~{\rm km}~{\rm s}^{-1}$ offset from 4C~23.56 can be
observed everywhere but the very edge of the detector.  The plot shows
that the emitters blueshifted by $\lesssim 1000~{\rm km}~{\rm s}^{-1}$ relative to
the radio galaxy will leave the filter at the center of the FOV, while
for  objects redshifted by $\gtrsim 1000~{\rm km}~{\rm s}^{-1}$ escape our detection
in the outer 1/4 of the detector. Though MOIRCS uses two independent
filters for the two channels, the difference of the center wavelength
of these two NBs is only 3~\AA~ ($\sim40~{\rm km}~{\rm s}^{-1}$ difference at
${\rm z}=2.49$). It should not cause any systematics between the channels.

We note that the velocity window of $\pm 500~{\rm km}~{\rm s}^{-1}$ is just adequate for
cluster member selection. Simulations of X-ray clusters shows that the
velocity dispersion of rich clusters generally gets smaller by about a 
factor of 2 as one goes from ${\rm z}=0$ to 3\ (e.g., \cite{eke98}).
Also, the previous H$\alpha$ emitter survey for a dense protocluster 
at ${\rm z}=2.16$ by \citet{kurk04b}
found a velocity dispersion of only $360~{\rm km}~{\rm s}^{-1}$ (narrower than the width
of the NB filter for their pre-selection), small for the inferred
richness of the protocluster. Thus we would expect to detect a
significant fraction of the protocluster H$\alpha$ emitters within the
MOIRCS narrowband window if its dynamical center is at the same
redshift and located on 4C~23.56.

\subsection{MOIRCS Data Reduction by MCSRED}

Data reduction of the MOIRCS science frames was performed using the
MCSRED package\footnote{Available at \protect\\
 http://www.naoj.org/staff/ichi/MCSRED/mcsred.html}, the
IRAF\footnote{http://iraf.noao.edu/} script pipeline for MOIRCS
imaging data developed by the author (IT). Here we briefly describe
the procedures in the pipeline for the potential user's convenience.

The first step is flatfielding the data.  We use a self-flat (i.e.,
maky a flatfield image from the dithered object images) for the
$K_{\rm s}$ band. An object mask is essential for making good quality
flatfields. To do this, we first make ``quasi-flatfielded'' images
by subtracting the adjacent image in time from the image from 
which we make  the object mask.  Such a 
``quasi-flatfielded'' image will show a flat sky and objects in the
image as positive-negative pairs.  Then the residual pattern in the
sky of each quasi-flatfielded image is further subtracted by fitting a
3rd-order polynomial surface. This procedure is done independently
for each quadrant, because MOIRCS is a 4-channel readout system and
any pattern related to the sky variation has discontinuities on
quadrant boundaries.  After replacing all the negative objects to the
sky value, cosmic-ray removal is applied. Each resulting image is then
smoothed by Gaussian kernel after a 3 by 3 boxcar median filter has been
applied. At this stage, most objects are clearly seen on the smoothed
image except very faint ones. Then any area that has a signal
larger than the threshold of $1.25\sigma$ of the original
pixel-to-pixel variation are replaced by the ``mask'' value of 1, with
an additional expansion to the area (here we set 5 pixels) for large
and bright objects. This quick object mask works well for the
usual science exposures. We note that, though we can do the more
accurate ``second-pass'' mask-making procedure after getting the final
combined image, the result is almost the same and is not necessary for
our data.

After making the object masks, the flat frame is made from the science
data itself with the mask applied. For the narrowband data we used
dome flat data as we describe further below. After the flatfielding of
the NB data, we further subtract the ``median sky'' image (made with
the object masks) which is made from the subset (typically 6 to 10) of
the frames before and after the image for the sky subtraction. 
For the $K_{\rm s}$ data this procedure can be skipped because the sky
pattern is relatively smooth (by the use of the self flat) and so we can 
subtract the sky in the next procedure reasonably well.
Then we further subtract any residual sky pattern remaining on each image 
using a 3rd-order polynomial surface fitting for each quadrant.

The resulting flatfielded, sky-subtracted images are then corrected to
remove the geometric distortion. The distortion correction database,
which is made from observations of astrometric reference
fields, is supplied by the Subaru Observatory. We then measure the position
of some (typically 20 to 50) bright and compact objects on each
image. The object catalogs for the images are further cross-matched
with the catalog for the first image in the input list, and from the
result we calculate the offset values we use to register the images to
the first image. Finally, these images are combined, where we weight by 
the rms$^{2}$ of the input images.

\vspace{1ex} A note on our NB2288 reduction: though we usually use the
sky- or self-flats for the reduction, we used the dome flat data for
correcting the pixel-to-pixel variation of the NB data. One reason is because the sky
counts in the narrow-band filter are generally low compared with that of
broad-band data. There is also the possibility that the sky may not be
flat in the NB filter. Due to the shift of the observable wavelength
window with position (see \S2.1.1), the inclusion or exclusion of OH
night-sky lines into the window becomes position-dependent.  Although
there is a small tilt pattern in the dome flats which is caused by the
difference of the light path between dome data and object data, we
estimate that the variation of the photometry due to the use of dome
flat is $<3\%$ ($\sim0.033$mag rms) by putting standard
stars at several positions across the field of view. This is small compared 
to our required accuracy.

\subsection{Photometry and the Catalog}
The object detection and the photometry are done using the
``double-image'' mode of SExtractor 2.5.0\ \citep{1996AA..117..393B}.
For the detection image, we used the un-smoothed version of the
narrowband data; therefore, our catalog is a simple narrowband-based
selection.  The $K_{\rm s}$ image is registered to this, and the
seeing size is closely matched (6.0 pixel, $\sim0.67$ arcsec FWHM) for
both images.  
The object detection is done within the area that is
covered by more than 12 frames ($75\%$) for NB2288 as well as 4 frames ($67\%$) 
for K$_{\rm s}$.  For the detection parameters, we use the SExtractor
  DETECT\_THRESH$=1.75$ and DETECT\_MINAREA$=15$ with a tophat
  detection filter (tophat\_4.0\_5x5.conv).

The effective survey area is 23.6 arcmin$^2$ which corresponds to a
survey volume of 840.7 comoving Mpc$^{3}$ at z=2.49 considering our NB2288
filter bandwidth. We used aperture magnitudes in a diameter of 12
  pixels ($=\timeform{1.4''}$ or 11.5 kpc physical scale) for color
measurement, and the MAG\_AUTO
values for derivation of the star-formation rates as well as for 
the stellar mass (\S3.5).

\vspace{1ex}

The standard star data for FS151 were reduced using the same
flatfielding data we used for the object reductions. The instrumental
magnitude for FS151 is measured from the individual 4 frames without
combining them because the images are slightly defocused.  These
magnitudes were then averaged to get the final measurement.  The
instrumental magnitude for each frame is measured by the
``growth-curve'' method which estimates the convergence magnitude with
increasing aperture radius. We used the IRAF APPHOT package.

The $K_{\rm s}$-band magnitude for FS151 is taken from
\citet{leggett06}, with the $+1.86$ conversion from the Vega to
AB-magnitude zeropoint applied \citep{tokunaga05}. For the NB2288
magnitude, we used the $K_{\rm s}$ magnitude and the low-dispersion
template spectra of a G-3 dwarf star by \citet{wh97}.
As a final check, we compared the $K$-band magnitude of some
unsaturated stars in our catalog to the 2MASS photometry. We find that
the magnitudes are well matched at the $\sim0.01$ magnitude level with
a scatter of $\sim0.1$.

\vspace{1ex}

Good photometric error estimates are critical for robust detection of
emission-line objects at lower equivalent widths (EWs). In our
analysis, we use the following two methods. First, we measure aperture
photometry in 1000 random (blank) regions to estimate the $1\sigma$
level of the sky variation. This yields 5-$\sigma$ limiting magnitudes
in a 12-pixel diameter aperture of 22.3 and 22.9 for $K_{\rm s}$ and
NB2288, respectively.  Second, we estimate the scatter of the $K_{\rm s}-$NB2288 color 
introduced by the photometric uncertainty.  We
generated 70000 fake objects with various magnitudes and placed
them randomly in blank image regions. After rebinning the sample
points in 0.5-magnitude intervals of NB magnitude, the scatter of the
output color excess ($K_{\rm s}-$NB2288) is evaluated by measuring the
distribution's $68\% (1\sigma)$, $95\% (2\sigma)$, and $99.7\% (3\sigma)$
levels directly from the simulation. The result of this simulation is
used for the estimate of the error on the color excess and on the
selection of emitters.

\subsection{Supplementary Data Used for Analysis}
In the current paper, we use some supplementary datasets (\S3.1 \&
\S3.4) and here we briefly describe them, though the full description
will be presented elsewhere (I. Tanaka et al. in prep.).

Deep $J$-band data were taken using the INGRID
camera\ \citep{packham03} on the William Herschel 4.2-m Telescope (WHT),
La Palma. The data were acquired on the photometric nights of UT 2001 Sep 7th
and 8th, under good seeing conditions ($\sim0.6$ arcsec.).  The
INGRID camera has a HAWAII HgCdTe $1024\times1024$ array, covering
roughly $4\arcmin\times4\arcmin$ field of view ($=\timeform{0.238''}$/pixel)
on the Cassegrain Focus of the WHT. The total integration time was
4800 s, resulting in a 50\% detection completeness limit of 24.4
mag. The data were reduced in the standard manner, similar to the
method for MOIRCS data reduction.

Deep $B$-band data were taken by the Suprime-cam on Subaru Telescope on
the evening of UT 2009 Nov 17th under clear but a marginal seeing
condition ($\timeform{0.9''}\sim \timeform{1.2''}$). It was taken as part of our
Ly$\alpha$ emitter search in this field. The net exposure time was
4500 s, having a limiting magnitude of 26.5 mag ($5\sigma$, \timeform{2.5''}-diameter aperture). The detail of the data reduction as well as the result of the
broadband SED fitting analysis are to be described elsewhere (I. Tanaka
et al., in preparation).

\vspace{1ex}

MOIRCS MOS spectroscopic observations of this field were done on UT 2006
September 7th (S06B-106: Kodama et al.); i.e., before the acquisition
of the narrow-band data we describe here.  Part of the target
selection was done using a similar narrowband survey executed in
September 2001 using the INGRID camera on WHT (IT and CP), which
should sample the redshift range of $2.429<{\rm z}<2.489$. The conditions
were clear, with medium ($\sim\timeform{0.8''}$) seeing conditions. We used
the grism with a spectral resolution of $\sim1000$ in the {\it K}-band
window covering the $2.0-2.4~\mu{\rm m}$ spectral range.  We used
$\timeform{0''.8}$ wide slits with typical slit lengths of
$\sim12''$. Each single exposure was 600~s to 1200~s, depending
on the airglow line strength. The standard A-B dither pattern was
applied during each exposure. In total 2-hrs of integration were
gathered. The MOS data reduction was done in the standard manner using
the MCSMDP pipeline (pre-release version) by one of the MOIRCS
builders (T.~Yoshikawa, private comm.).  Details of the MOS reduction
procedure by MCSMDP can be found in the recent paper of
\citet{yoshikawa10}.

\vspace{1ex}

Finally, the deep 24 $\mu$m imaging from {\it Spitzer}/MIPS were obtained as
part of GO program 30240 (PI: A. Stockton). The total exposure time in
the central $2\arcmin\times2\arcmin$ of the mosaic is approximately 
5 ks. The 171 separate exposures each had a background fitted and then 
subtracted before a median of the array was determined. This median was 
then subtracted from each exposure. The exposures were then mosaicked 
with MOPEX software\footnote{http://ssc.spitzer.caltech.edu/dataanalysistools/tools/mopex/} with standard inputs and resampling the pixels by a factor of 2.

For the current analysis we focus on a subset of the 24 $\mu$m-selected sources 
that have clear IRAC counterparts.
The MIPS photometry was extracted using DAOPHOT in IRAF to fit an
empirical PSF model to multiple proximate (within $12 ''$)
sources simultaneously.  The MIPS source positions were pre-defined
using the position of the closest IRAC counterpart.  This photometry
method mitigates the effects of source confusion.

\section{Results}

\subsection{Emission-line Object Selection}

\begin{figure*}
\begin{center}
\FigureFile(160mm,100mm){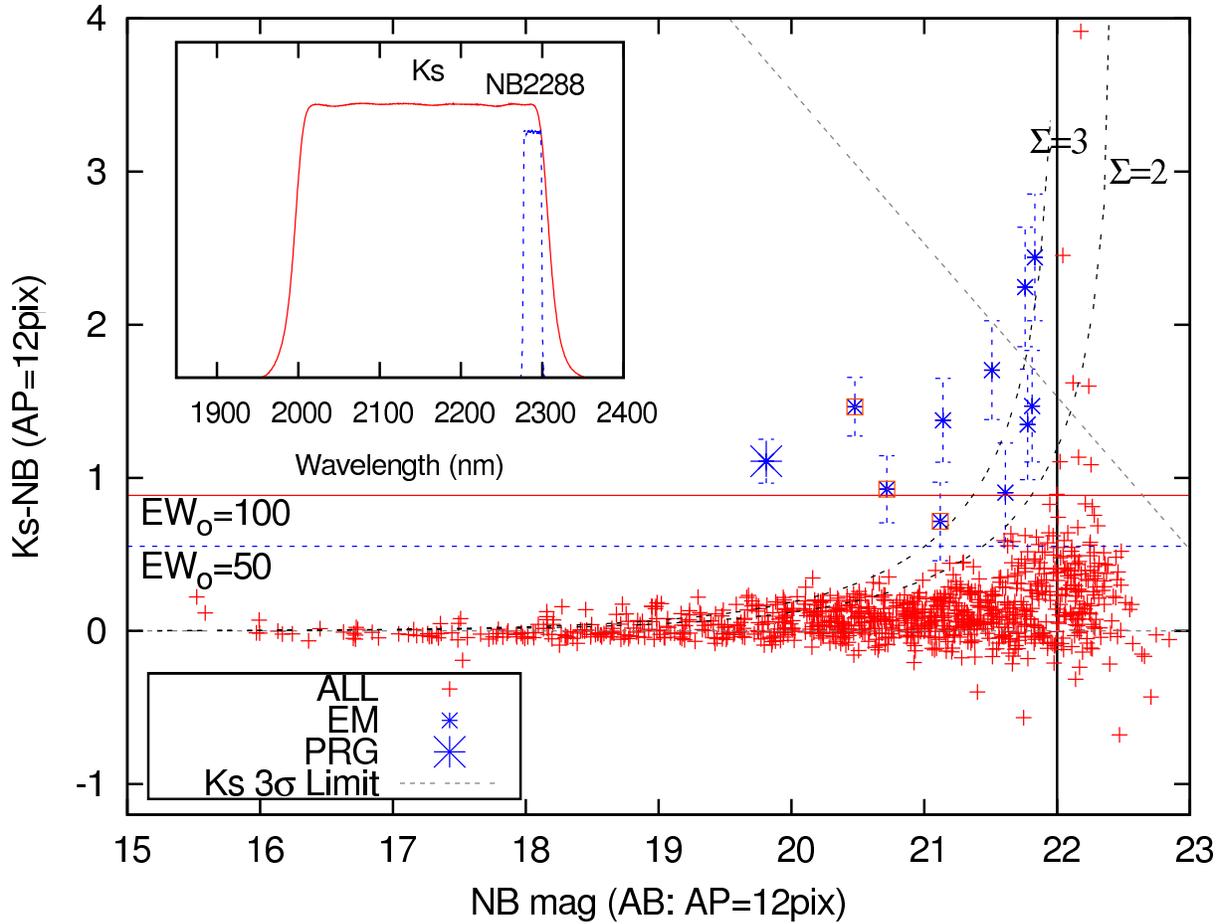}
\end{center}
\caption{The $K_{\rm s}-$NB versus NB color-magnitude diagram for our cataloged objects (plus with the label ``ALL''). The blue asterisks (labeled as ``EM'') indicate our selected emission-line objects based on the criteria EW$_{o} > 50$~\AA, the narrowband aperture magnitude of less than 22 mag, and $\Sigma > 2$, which is used for analysis in the paper (one spurious object in the sample is removed here). The brightest emitter is 4C~23.56 which is shown as a large asterisk. Emitters with red open boxes are the objects with MOIRCS spectra (except 4C~23.56). The tilted 4-dashed line indicates the 3-sigma limiting magnitude of $K_{\rm s}=23.5$. The insetted figure shows the transmission curve of the filters used in this study. \label{fig2}}
\end{figure*}

The selection of H$\alpha$ emitters at ${\rm z}\sim2.5$ is based on excess
emission in NB2288 measured via $K_{\rm s}-$NB2288 color.  In
Figure~\ref{fig2}, we show the color-magnitude diagram for all objects
detected in our observations. The overall distribution of the detected
objects tends to become broader with increasing magnitudes due to the
increasing photometric uncertainty. The intrinsic scatter is also seen
for the brighter objects. The intrinsic scatter is caused by the
variation of the spectral energy distributions (SEDs) in the $K_{\rm s}$
band.
%
%

The two curves labeled as $\Sigma=2$ \& 3 in Fig.\ref{fig2} are the
conventionally used measures of the significance of narrowband
excess (e.g., \cite{geach08}). It is formulated as below,
\begin{equation}
m_{\rm K}-m_{NB}=-2.5 log [1-\Sigma\delta 10^{-0.4(m_{z}-m_{NB})}],
\end{equation}
where $m_{z}$ is the zeropoint of the narrowband image and $\delta$ is
the photometric uncertainty. The $\Sigma=2$ line generally matches
well with our simulated $\sigma=2$ curve.
As the color scatter becomes large at NB2288$>22$ mag, we limit the
selection of emission-line objects to $\Sigma>2$ and NB2288$\leq22$
(corresponding to $7\sigma$ level in 12-pixel aperture). We note that
the use of our simulated $\sigma=2$ scattering curve does not change
the sample.  At NB2288$> 22$, there are actually only a few
emission-line object candidates with $\Sigma>2$, due to the steep rise
of the selection curve.

At the bright end, the intrinsic scatter is relatively large due to
the variation of the SEDs in the $K_{\rm s}$ band window and the
selection of emitters with low-equivalent width may introduce 
spurious sources. We therefore introduce a further selection criterion
using the rest-frame equivalent width, EW$_{o}>50$\AA\ at
${\rm z}=2.49$. This corresponds to the $K_{\rm s} - $ NB2288 $ = 0.55$ mag.
The limiting H$\alpha$ line flux of the emitter sample by these
criteria is roughly $7.5 \times 10^{-17} {\rm erg}\ {\rm s}^{-1}\ {\rm
  cm}^{-2}$. These values are computed using the same formula used in
\citet{geach08}.

We find 12 objects that satisfy the selection criteria set
here. Removing one object which is found to be a diffraction spike
by visual inspection, the remaining 11 objects are cataloged as robust
emission-line objects for further analysis. We note that the brightest
object in the emitter sample is 4C~23.56.  The observational
properties of the emitter sample are summarized in Table~\ref{tbl1}.
%
%

\vspace{1ex}
\subsubsection{Spectroscopy}
\begin{figure*}
\begin{center}
\FigureFile(160mm,100mm){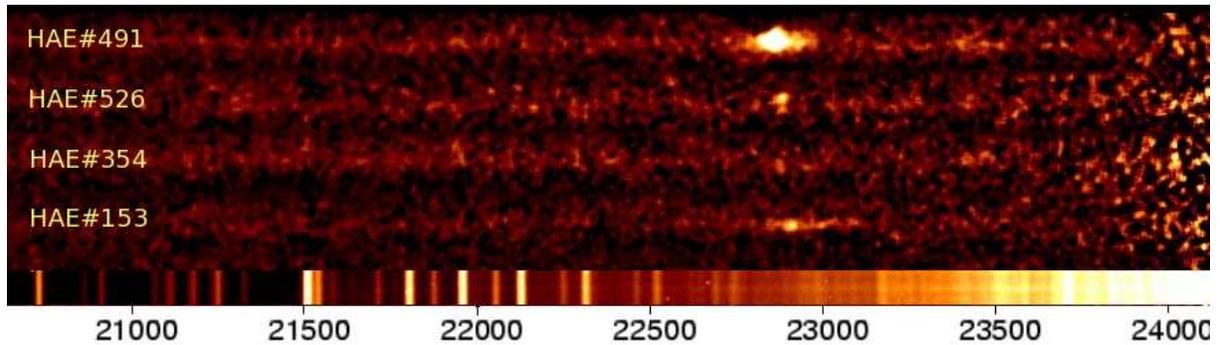}
\end{center}
\vspace{3ex}
\caption{MOIRCS MOS spectra of four H$\alpha$ emitters. The top spectra (HAE\#491) is 4C~23.56. The barcode-like pattern at the bottom is the OH airgrow spectra. 
The numbers at the bottom are wavelengths in angstrom.\label{fig4}}
\end{figure*}

\begin{figure*}
\begin{center}
\FigureFile(60mm,100mm){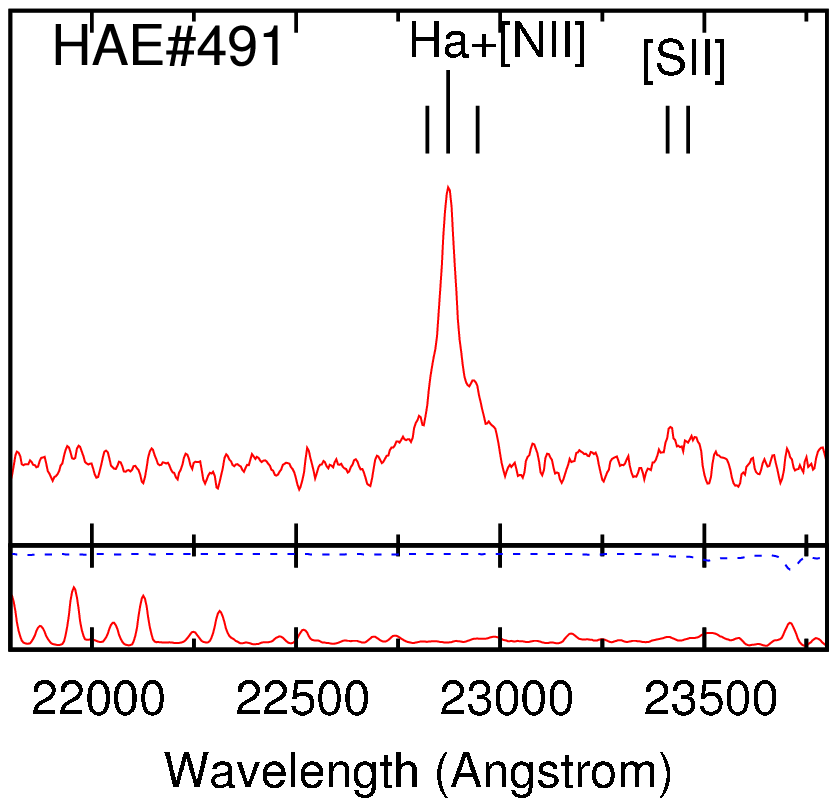}
\FigureFile(60mm,100mm){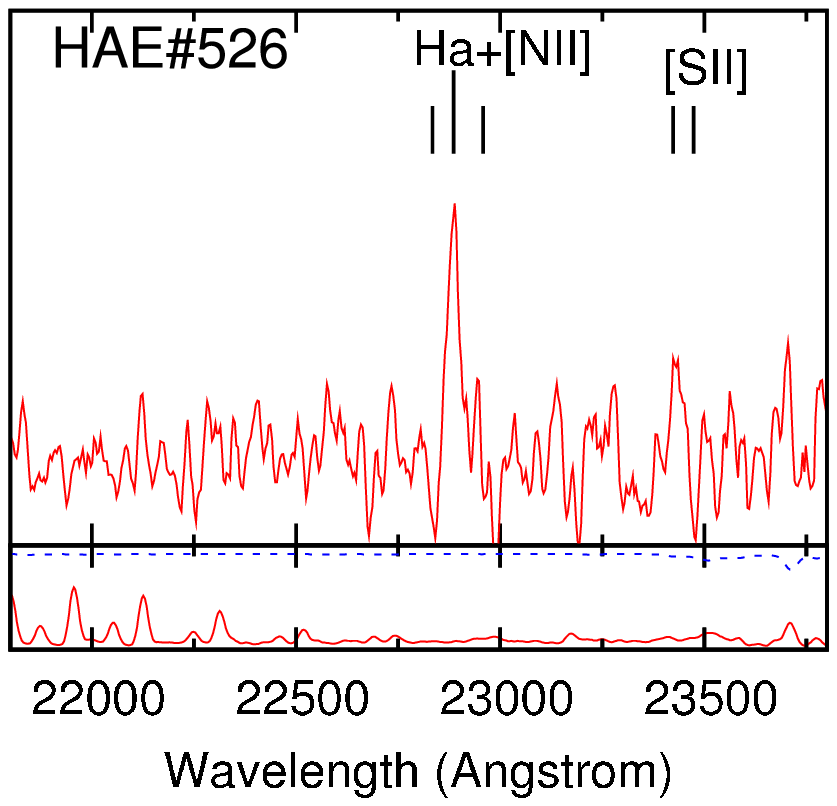}

\vspace{5ex}
\FigureFile(60mm,100mm){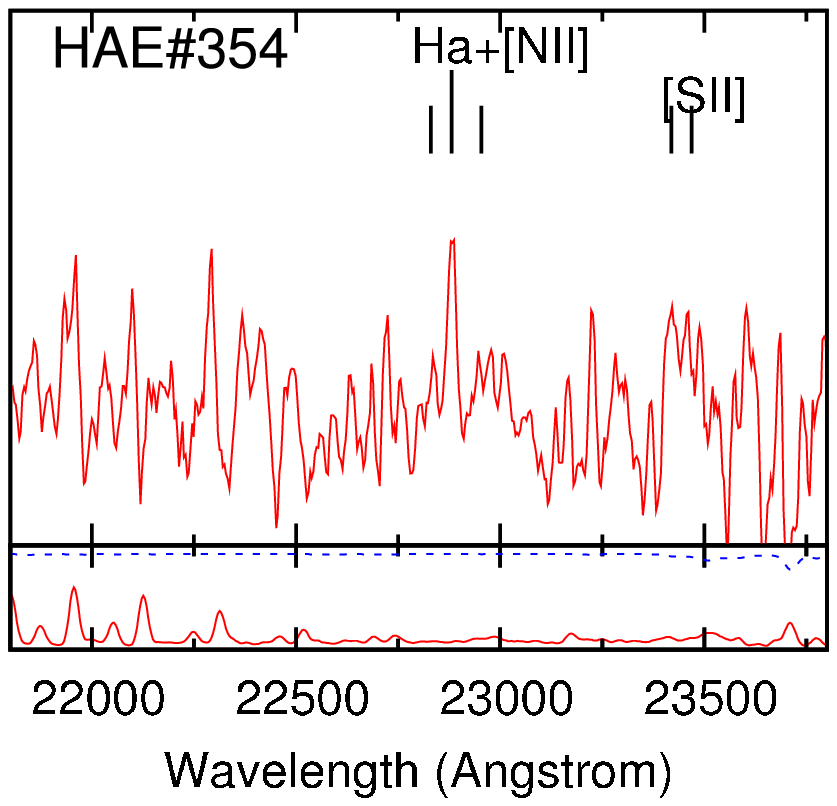}
\FigureFile(60mm,100mm){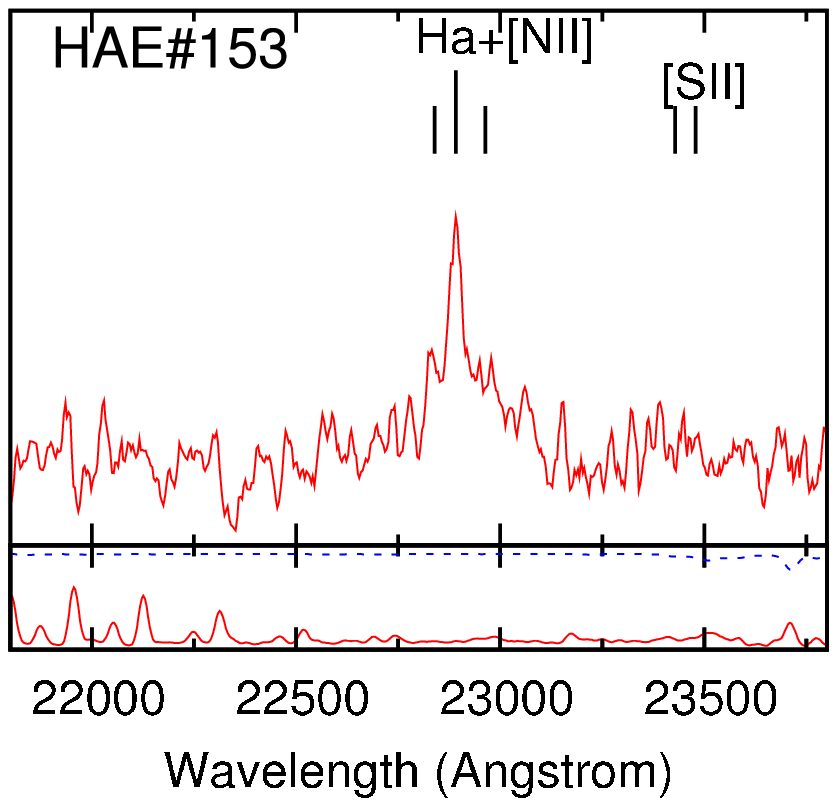}
\end{center}
\vspace{3ex}
\caption{MOIRCS MOS spectra of four H$\alpha$ emitters. The top left spectra (HAE\#491) is 4C~23.56. The OH airglow lines (red solid line) and the atmospheric transmission curve (blue dotted) are also plotted at the bottom panel of each spectra. Note that the spectra are not calibrated.\label{fig5}}
\end{figure*}

Though spectroscopic observations for all the emitter candidates are
not obtained yet, four of the 11 emission-line candidates have NIR
spectroscopy.  These data were taken before the acquisition of the
current NB data. The one- and two-dimensional spectra for each target
are shown in Figures~\ref{fig4} and \ref{fig5}.  HAE\#491 is the radio
galaxy 4C~23.56 itself.
%
%

Due to the relatively low signal-to-noise ratio (SNR) of the
spectroscopy, unambiguous identification of H$\alpha$ via detection of
the nearby [N{\sc ii}] or [S{\sc ii}] lines is only possible for the radio
galaxy. The objects HAE\#526 and HAE\#153 are relatively bright, and
we can say this is not [O\emissiontype{III}]$\lambda 5007$ emission at ${\rm z}=3.571$ because of the
lack of the [O\emissiontype{III}]$\lambda 4959$ near 2.267$\mu$m (as well as the lack
of H$\beta$ at $2.222~\mu$m).  Furthermore, the spectra of the latter
object seems to have a hint of the broad component which would not
appear if it is [O\emissiontype{III}]. The lines from these two sources may also be
from the Paschen series. If it is the Pa$\alpha$ (Pa$\beta$) line,
Pa$\beta$ (Pa$\gamma$) will be observed at 1.56 (1.95) $\mu$m. Unfortunately our 
spectra do not cover wavelengths below $2~\mu$m.  We
have to make use of other information to exclude the latter possibility
(see next section).  For the object \#354, the H$\alpha$
  signal at $2.288~\mu$m is quite uncertain.  However, we see a
  possible signal at $2.342~\mu$m at the expected position of the [S\emissiontype{II}]
  line. Though both lines are quite low SNR, the inferred large
  [S\emissiontype{II}]/(H$\alpha$+[N\emissiontype{II}]) ratio of $\sim1$ would indicate that the
  object harbours an AGN \citep{osterbrock89}.  We list the measured
redshift values in Table~\ref{tbl1} supposing the line is H$\alpha$.

The velocity difference among these emitters is relatively small. They
lie in the range of $2.287-2.289~\mu$m (230~km~s$^{-1}$ range), which is in
the red edge of the INGRID narrowband filter function we used for
pre-selection ($2.25-2.29~\mu$m). If these emission lines are not
H$\alpha$ but from the interlopers unrelated to 4C~23.56, we can
expect a more scattered distribution in wavelength range across the
INGRID filter function. This is a strong indication that they are
physically associated with each other and 4C~23.56.

\vspace{1ex}

\subsubsection{Broadband Colors of HAEs}
Our emission-line survey for H$\alpha$ at ${\rm z}=2.49$ may contain
interlopers at different redshifts. As we showed in the previous
section, the spectroscopic detection of a single line remains
ambiguous; it could be Pa$\alpha$ for example.  \citet{geach08} used
color selection as well as photometric redshifts to distinguish
interlopers in their H$\alpha$ emitter sample at ${\rm z}=2.2$. They
estimated that, among their 55 ``robust'' sample galaxies, a third can
be lines other than H$\alpha$ (12 Pa$\alpha$, 6 Pa$\beta$, and
one [O\emissiontype{III}]). Their result indicates that foreground objects
(Pa$\alpha$ objects at ${\rm z}=0.22$ and Pa$\beta$ objects at ${\rm z}=0.78)$
could be the major contaminants of our data.

To make sure that our detected emission-line objects are mainly
H$\alpha$ emitters, we measure the continuum color of our
emission-line candidates using our supplementary Subaru $B$-band and
WHT $J$-band data (\S2.4).  The $J$-band data covers the central
$\timeform{3.8'}\times\timeform{3.4'}$ area of the MOIRCS field of view (here we
refer to the area as the ``INGRID subfield'' as shown in
Figure~\ref{fig6}), while the whole area is covered by $B$-band data.

Though our selected emission-line candidates all lie in the INGRID
subfield, 4 objects are not seen in our $B$-band data. Two more
objects are overwhelmed by the PSF of bright nearby stars. We could
measure a $BJK$ color for five objects. Fortunately, all four
spectroscopic targets are in this sub-sample. The $B-J$ versus
$J-K_{\rm s}$ two-color diagram is shown in Figure~\ref{fig3}a. The
H$\alpha$ emitter candidates are marked by filled boxes (red).

As discussed in detail in the Appendix,
the $B-J$ versus $J-K_{\rm s}$ two-color diagram is found to be quite
effective for rejecting objects at ${\rm z}<1$. Figure~\ref{fig3}b is the
$BJK$ two-color distribution of galaxies in GOODS-North region from
the MOIRCS Deep Survey (\cite{kajisawa09, kajisawa10}, and references
therein). The objects that satisfy $(J-K_{\rm s}) >
0.4~(B-J)-0.2$ are empirically shown to have spectroscopic redshift of
${\rm z} > 1$ from Fig.\ref{fig3}b.
%
%

The position of our 5 emission-line objects in Fig.\ref{fig3}a is
quite favorable; with the exception of one object, they satisfy the
$BJK$ criterion.  These 4 $BJK$ emitters comprise our
spectroscopic sample. Thus we can conclude that the emission line we
detect spectroscopically is indeed H$\alpha$, not from the Paschen
series. This is strong support for the existence of a protocluster at
${\rm z}=2.5$ in the 4C~23.56 field.

One object whose color is below the $BJK$ line could be an interloper,
though we cannot conclude this for certain until its spectra is obtained. One such
object can be expected from the field survey as we see in \S4.1.
During the following discussion we consider that all our selected
emission-line objects are HAEs at ${\rm z}\sim2.49$,
though a few of them might be foreground objects.

\subsection{Spatial Distribution of HAEs}

The spatial distribution of our HAE sample on the sky is displayed in
Figure~\ref{fig6}. The position of 4C~23.56 is marked with the large
diamond. The distribution of our selected HAEs (shown with open red
circles) is quite tightly clustered at the east to south-eastern side
of the radio galaxy.
%
%

The biased distribution of HAEs to the eastern (i.e., channel-1) side
of 4C~23.56 is not due to the wavelength shift of the filter
transmission window or any difference between
channels.  
It is also not due to some unexpected trouble in the channel-2 filter,
because 4C~23.56 and the adjacent two emitters are actually detected
on the channel-2 side. The inhomogeneous distribution seems to be the
reflection of the real distribution of the emitters.

If we suppose that the majority of the protocluster members are
strongly redshifted relative to 4C~23.56, the situation could
change. As is discussed by \S2.1.1, HAEs redshifted by $>$1000~km/s
relative to 4C~23.56 will drop off from our detection window in the
outer half of the FOV. In the inner half of the FOV we should detect
objects with large velocity difference. Though our three HAEs with
spectroscopic redshift have small velocity offset ($<230~{\rm km}~{\rm s}^{-1}$), further
spectroscopy for all the HAEs may uncover a biased velocity distribution.

\subsection{Offset H$\alpha$ Emission in the ``Eastern Clump''}

Here we comment on the objects in the area enclosed by the small white
box in Fig.\ref{fig6}. There is a group of 4 emitters that lies within
a $15''$-diameter area (=300 kpc comoving), which is also shown
in Figure~\ref{fig7} as a zoomed view.  We call this area the
``Eastern Clump''. The compact size of the HAE group suggests that
they might eventually merge into one massive galaxy in the future and
become (one of) the brightest cluster member(s).
%
%

One notable character of HAEs in the Eastern Clump is that the light
distribution of each HAE in the NB2288 image is offset from the
$K$-band object position, as shown in Fig.\ref{fig7}. Especially
interesting is that the offset direction of emission-line gas in three
HAEs on the west side of the clump in Fig.\ref{fig7} is roughly
aligned to the North to North West.
%
%

To show these offsets more clearly, we ran SExtractor on each image
independently and measured the positional offset of objects detected
in both images. The result is shown in Figure~\ref{fig8}. In the
figure, the 1-, 2-, and 3-$\sigma$ scatter, based on the objects that
have similar magnitudes to the HAEs (dots), are shown by
circles of dashed lines. The HAEs are marked as filled squares,
with the objects in the Eastern Clump shown in Fig.\ref{fig7} overlaid
with asterisks. 
There are two more large-offset ($>3\sigma$) objects seen in
Fig.\ref{fig8}. One of them is indeed the object just south of the
eastern clump. We confirm that the offset of the emission-line gas is
not due to poor astrometric alignment of the two images; all the other
objects adjacent to the Eastern Clump show the same scatter shown as
dots here.

The $K$-band light centroid can be considered the center-of-mass for
each galaxy.  Therefore, these offsets of the H$\alpha$ emission
indicate that the star-formation is occurring at the periphery of the
galaxies. This could be due to interactions among the galaxies
triggering star-formation (e.g., \cite{sulentic01, cortese06}),
though the apparent alignment of the offset direction is difficult to
explain. The apparent offset of the H$\alpha$ emission is roughly
$\timeform{0''.2}-\timeform{0''.4}$, corresponding to a physical size of
$1.6-3.2$~kpc at ${\rm z}=2.49$.

Another possible explanation may be that vast outflows of
emission-line gas are being launched from these galaxies.  The
direction of such outflows could be influenced by the direction that
the group is moving towards the center of the cluster potential. If
the cluster potential is already filled by dense intracluster gas, the
expelled emission-line gas from these galaxies may leave a ``wake''
behind each galaxy. In the local universe, similarly offset light from
emission-line gas is caused by ram-pressure stripping of galaxies
by intracluster gas (e.g., \cite{yoshida02, yagi07, sun07}). For
example, \citet{cortese07} discovered a trail of blue emission-line
knots behind an object in Abell2667 at ${\rm z}=0.23$. Our detected offset
of the H$\alpha$ emission in the Eastern Clump may be similar.

A detection of extended X-ray emission from 4C~23.56 intracluster gas
is the key for falsifying the ram-pressure stripping scenario. The 
field was observed by {\it XMM} and reported by \citet{johnson07}. Though
they detected extended soft X-ray emission near the radio galaxy
in a $\sim30$ ksec EPIC exposures, they concluded that 
inverse-Compton scattering of CMB photons is the most likely emission
mechanism. Indeed the position and direction of the X-ray
extension is well aligned with the radio lobe which extends $\sim0.5$
Mpc north-east/south-west from the radio galaxy, and not with the
direction where the Eastern Clump lies. For now, there is no evidence
of thermal X-ray emission from hot intracluster gas.

\subsection{HAEs and the {\it Spitzer} MIPS Data}

{\it Spitzer} MIPS $24~\mu{\rm m}$ imaging has been used to probe 
star formation activity at z$\sim$2 (e.g., \cite{daddi07}).  There
is public, archived deep $24~\mu{\rm m}$ MIPS data for the 4C~23.56 field
(\S2.4).  If the protocluster around 4C~23.56 is characterized by
active star-formation, we may also see it as an excess of faint MIPS
sources.

In Figure~\ref{fig13} we show our re-processed MIPS $24~\mu{\rm m}$
image. The strong point source at the center of the image is
4C~23.56. We can clearly see an excess of faint MIPS sources around
the radio galaxy compared to the outer parts of the image. 
We also find that 9 objects out of our 11 HAE sample
(shown as red open circles) have MIPS counterparts.  However, the
apparent excess of $24~\mu{\rm m}$ sources does not seem to be due solely
to the HAE counterparts. Here we examine the excess of 
MIPS-selected objects in the 4C~23.56 field. The $24~\mu{\rm m}$ photometry
and the inferred star-formation rates for the HAE counterparts
are separately discussed in the next section (\S3.5).
%
%

To estimate the density excess of MIPS sources quantitatively, we have
to consider the difference of the exposure coverage across the field.
We first exclude any area that has exposure time less than $1/3$ of
the total integration. We then set the SExtractor object detection
threshold to $2.8\sigma$. This is higher than usual so that the
detection is not affected by the S/N difference due to the variation
of exposure time across the field. 
To check the robustness of the detection threshold further, we 
reversed the pixel count of the MIPS image and executed the object
detection with the same SExtractor parameters. Besides a clear pixel
defect, we detected no sources in the inverse image, indicating that the no fake
object from the sky variation is included in the current analysis.
Furthermore, we exclude all objects with flux density less than $90~\mu {\rm Jy}$, 
as they may suffer from the source confusion in crowded regions.

We also introduce an upper flux density cut of $275~\mu {\rm Jy}$ because most of
our HAEs are fainter than $200~\mu {\rm Jy}$. Any object much brighter
than $200~\mu {\rm Jy}$ has a high possibility of being in the
foreground.  The upper flux density cut is chosen to be roughly twice the
median $24~\mu{\rm m}$ flux density of our HAEs.

Note that we apply the SExtractor detection directly to the $24~\mu{\rm m}$
data for the source density excess estimate.  For the SFR estimate
from the MIPS photometry for our HAE sample we use DAOPHOT software as
described in \S2.4, because it provides a more robust flux estimate of
each source.

The resulting density map of the faint ($90 - 275~\mu$Jy) MIPS
sources is overlaid on Fig.\ref{fig13} as contours.  The distribution
of sources has been smoothed by a Gaussian kernel of
$\sigma=30''$. For reference, the position of the HAEs are shown
as red open circles. It is quite clear that the MIPS source density
excess closely matches the HAE distribution.

To quantify the density excess of MIPS sources further, we define
the ``excess region" as the $80''$-radius region around the peak
density contour, and compare the excess within the region with the
``outer region'' at larger radii. From the density of faint MIPS
sources in the ``outer region'' we estimate the expected number in the
``excess region'' as 15.3. On the other hand, we count 39 objects in
the ``excess region'', yielding a number enhancement of
$2.55^{+0.59}_{-0.48}$. This corresponds to a 4.5-$\sigma$ excess by
proper Poisson statistics \citep{gehrels86}. As a crosscheck, we downloaded
the deep MIPS data for GOODS-N/S fields\footnote{available in 
http://ssc.spitzer.caltech.edu/spitzermission/observingprograms/legacy/goods/} and 
counted sources in the same 
manner, yielding the expected count of $18.2\pm0.7$ for an area with the size of the 
"excess region". The source counts of deep MIPS data for ELAIS-N1 field by \citet{chary04} also shows
the expected number of 17.0. Our number is consistent with those surveys, considering
the uncertainty by the cosmic variation.

The HAEs with MIPS
counterparts (9) can only explain less than half of the total excess
count ($\sim24$). This implies that our shallow H$\alpha$ emitter
survey may be missing an even larger number of the dust-enshrouded
star-forming objects in the excess region. Future follow-up observations 
using ALMA etc would be quite interesting to understand these hidden dusty sources.

\subsection{Star-Formation Rate and Stellar Mass of HAEs}

In this section we estimate the star-formation rate (SFR) and the stellar mass of our HAE sample. We note, however, that the estimation of the SFRs and the stellar masses for our HAE sample are uncertain and should be treated with caution.

\vspace{1ex}
\subsubsection{H$\alpha$-based Star-Formation Rate of HAEs}

It is known that dust absorption lowers the measured SFR from
H$\alpha$ emission relative to the intrinsic SFR based on radio or FIR emission 
(e.g., \cite{hopkins01, afonso03, erb06, daddi07}). The
contamination of the [N\emissiontype{II}] 6583\AA\ emission line to the narrowband
flux is also an unknown factor (here we set the factor as 0.2). The
contribution of AGN to the ionizing radiation also adds 
uncertainty to the estimate. These uncertainties can be calibrated
using future deep spectroscopy of complete HAE samples.

\citet{kennicutt98} gives the conversion formula from $L(H\alpha)$ to
the SFR as follows.
\begin{equation}
SFR~({\rm M}_{\solar}~{\rm yr}^{-1})=7.9\times 10^{-42}~L(H\alpha)~{\rm (erg~s^{-1})}
\end{equation}
Though the conversion assumes the standard \citet{salpeter55} Initial
Mass Function (IMF), the IMF is known to overestimate the stellar
mass. We here use the conversion correction for the \citet{chabrier03}
IMF by dividing the value by 1.7, for mainly practical reasons, so
that our data can be compared to some recent H$\alpha$-based SFR
estimates in the literature.  
Note that the dust extinction should affect the SFR estimate much more
than the uncertainty due to [N\emissiontype{II}] contamination. An AGN 
contribution can also exist for some galaxies; the unusually high derived SFR for 4C~23.56 is clearly due to the AGN contribution to the H$\alpha$ flux.

As for the extinction-correction factor, the $BJK$ diagram
(Fig.\ref{fig3}a) is a useful diagnostic for a part of our sample. In
the figure we show the A$_{V}=1$ extinction vector using the
\citet{calzetti01} extinction law.  Assuming that the objects are pure
disks with dust extinction, the color distribution of the HAEs
indicates that the level of the extinction ranges from 0 to $\sim
1$. Using the \citet{calzetti01} recipe and the conversion of
A$_{H\alpha}=0.82 $A$_{V}$, an extinction correction factor of 1 to 5
is estimated.

For the remaining 6 objects we cannot estimate the extinction.  For
comparison, the recent study by \citet{onodera10} estimated the level
of extinction for their $sBzK$-selected H$\alpha$ emission-line
objects that seem to have roughly similar stellar masses as our HAEs
(\S3.5.3). Their extinction estimate is based SED
fitting with additional correction as prescribed by
\citet{cidfernandes05} and \citet{savaglio05}. Their Table~4 indicates
that the average of the extinction correction factor is 4.9. We
applied the correction factors of 4.9 for the objects we could not use
the $BJK$ diagram. The raw and the extinction-corrected SFR for each
HAE is listed in Table~\ref{tbl2}.
%
%

\vspace{1ex}
\subsubsection{$Spitzer$ MIPS Star-Formation Rates}

The bolometric infrared luminosity of galaxies correlates very well
with other measures of the star-formation rate and may be the most
robust measurement of the SFR.  The infrared light is emitted by
dust-reprocessing of the emitted UV light from young stars and
therefore peaks at rest-frame wavelengths between 100 and 300~$\mu$m.  In
lieu of observations at these frequencies, the longest wavelength
datapoints need to be extrapolated using assumed SED models.  For this
radio galaxy field we have used the observed $24~\mu$m imaging to infer
the SFRs for the discovered H$\alpha$ emitters.

We used the SED template for Arp 220, a possible low-redshift
analogue, to correct the $24~\mu$m to rest-frame $8.0~\mu$m
luminosities \citep{rieke09}.  From these values we estimated the
bolometric infrared luminosity using the \citet{CE01} and \citet{DH02} 
SED models, following the prescription by \citet{daddi07}\footnote{The
formula (4) in \citet{daddi07} has a possible typo. The constant should be +1.745,
not +1.23, to match the desciption in the text and to reproduce the relation in their Fig.7}. The resulted infrared luminosities
from the two SED models are averaged.
Finally, we use the \citet{kennicutt98} $L_{\rm IR}$ to SFR conversion below,
\begin{equation}
SFR~({\rm M}_{\solar}~{\rm yr}^{-1})=3.0\times 10^{-44}~L_{\rm FIR}~{\rm (erg~s^{-1})}.
\end{equation}
Note that the original formula is divided by 1.5 for \citet{chabrier03} IMF \citep{yoshikawa10}.
These conversions and extrapolation result in SFRs which are uncertain by up
to factors of a few \citep{papovich07}.  For our purposes, to estimate
the FIR output from likely protocluster galaxies, these uncertainties
are tolerable.  The {\it Herschel Space Telescope} has already
quantified these uncertainties for other fields (e.g., \cite{rodighiero10})
and may eventually provide longer wavelength data for these and other
protocluster galaxies. The result of the MIPS-based SFR is also listed
in Table~\ref{tbl2}.

The resulting MIPS-based SFRs are significantly (a factor 2-14 with median of 3.6) higher
than the SFRs based on extinciton-corrected H$\alpha$. We will discuss this further in \S4.3.

\vspace{1ex}
\subsubsection{Stellar Mass of HAEs}

Next we estimate the stellar mass of our HAE emitters. We employ the
conversion formula for the stellar mass from the $K_{\rm s}$-band
photometry by \citet{daddi04}. They showed that for galaxies at
$1.4<{\rm z}<2.5$ the stellar mass is reasonably estimated using the formula
below,
\begin{eqnarray}
\log ({\rm M}_{*}/10^{11}~{\rm M}_{\solar}) = &&-0.4~(K^{\rm tot}_{\rm Vega}-K_{11}) \nonumber\\
&& + 0.218~[(z-K)_{\rm AB}-2.29]
\label{eq4}
\end{eqnarray}
where $K_{11}$ is the $K$-band magnitude (Vega scale) of the average
$10^{11}~{\rm M}_{\solar}$-mass objects of their sample (=20.14 for the average
redshift of $\langle {\rm z}\rangle =1.9$). Considering the difference of the luminosity
distance between $\langle {\rm z}\rangle =1.9$ and 2.49, and the average distribution of
$(z-K)_{\rm AB}$ color of 1 from the Fig.7 of \citet{daddi04}, we derive
the estimate of the stellar mass at ${\rm z}=2.49$ as
\begin{equation}
\log ({\rm M}_{*}/10^{11}~{\rm M}_{\solar})_{2.49}~=~-0.4~K^{\rm tot}_{\rm AB}+8.8
\label{eq5}
\end{equation}
where $K^{\rm tot}_{\rm AB}$ is in AB-magnitude scale.
%
%

As a check of the validity of the formula, we compare it to the
stellar mass estimates based on SED fitting the 12-band photometry for
the GOODS-North MOIRCS Deep Survey Catalog (for details of
the model fitting, see \cite{kajisawa09, kajisawa10}). 
Figure~\ref{fig10} shows
the stellar mass versus $K$-band magnitude of the cataloged objects in
the redshift range $2.3<{\rm z}<2.6$, for objects with photometric redshifts (open
boxes) and spectroscopic redshifts (crosses). Though the Daddi estimator by
equation~(\ref{eq4}) fits the overall distribution of the photometric redshift sample
reasonably well, the slope seems a bit shallower than the
data. Therefore we fit all the photo-z-based data in Fig.\ref{fig10}
and got the best fit line as
\begin{equation}
\log ({\rm M}_{*}/10^{11}~{\rm M}_{\solar})_{2.49}~=~-0.5127~K^{\rm tot}_{\rm AB}+11.312.
\label{eq6}
\end{equation}
We use this formula for the conversion from $K$-band magnitude to
total stellar mass. The rms scatter of the fit is calculated and found
to be only 0.2~dex, consistent with the value quoted by
\citet{daddi04} (but note that the scatter increases to 2.4~dex if we
use brighter subsample of $K_{\rm s}<23$).  The resulting stellar mass 
for each HAE galaxy is
listed in Table~\ref{tbl2}. We note that the $K_{\rm s}$-band
magnitudes for HAEs will be affected by the H$\alpha$ emission. The
rest-frame equivalent width of 100~\AA\ will affect the mass estimate
by 11\%. The median value of the offset due to line emission is
$\sim 20\%$. This is small compared to the mass estimation
uncertainty of $>0.2$~dex. We also calculate the
specific star-formation rate (SSFR) for the HAEs. The results are shown 
in Table~\ref{tbl2}.

\section{Discussion}

\subsection{Comparison to Field Counts}

Our discovery of 11 HAE candidates to a limiting flux of $7.5 \times
10^{-17}$\ erg\ s$^{-1}\ $cm$^{-2}$ in 23.6 arcmin$^{2}$ area should be
compared with the number expected in the general
field. \citet{geach08} provide us with excellent reference data. Their
HiZELS survey measured the number density of H$\alpha$ emitters at
${\rm z}=2.23$ over 0.6 deg$^{2}$ using the narrow-band filter (H$_{2}$S1,
$\lambda_{c}=2.121~\mu$m, $\delta\lambda=0.021~\mu$m) to a line flux
limit of $10^{-16}$\ erg\ s$^{-1}\ $cm$^{-2}$. Their survey field is 91.5
times wider than our own and their filter has a similar bandwidth to
our NB2288 filter. They detected 180 emitters including non-H$\alpha$
(e.g., Paschen series, [Fe\emissiontype{II}], [O\emissiontype{III}]) objects under their selection
criteria of {\it observed} EW of 50~\AA\ ($=14.3$~\AA\ in rest frame at
${\rm z}=2.5$). As our data is deeper than theirs, we set our detection
limit to match theirs.  To a NB magnitude of 21.3 mag we find 5
emitters. Next we have to estimate the how many of their 180 emitters
would satisfy our threshold of {\it rest-frame} EW of 50~\AA. In the
filter bandwidth they used, the rest-frame EW$_{o}$ of 50~\AA\ (at
${\rm z}=2.49$) corresponds to a narrowband excess of 0.56. We estimate
the number of emitters that satisfy the EW$_{o}>50$~\AA\ to be about 90
from their Figure~1. Scaling the number to match our field of view, we
derive the expected number of the {\it field} emitters in our field to
be $\sim 1$.
Note that, similar to our sample of HAE candidates, this field 
count includes interlopers too.
Considering our total of 5 emitters, we can say that our field contains 
5 times the expected field count.

Recently \citet{geach10} have published more generalized HAE counts from their 
HiZELS and other surveys. Their Table 2 gives the expected field HAE number 
counts at the limit of $1~10^{-16}$\ erg\ s$^{-1}\ $cm$^{-2}$ (EW$_{o}>10$~\AA)
to $\rm{z}=2.2$. Extrapolating their result to $\rm{z}=2.5$ by third-order polynominal
and applying our survey size, we get the exptected field count in our survey area as 
$1.2^{+0.63}_{-0.46}$. If we again suppose roughly half of them can be observed as HAE in 
our EW cut of EW$_{o}>50$~\AA, the expected field count could be only 0.6. Thus the result suggests the overdensity even more strongly.

A deep and wide narrowband HAE survey using the new instrument HAWK-I was recently
published \citep{hayes10}.  
A field of $\timeform{7.5'}\times\timeform{7.5'}$ in GOODS South was surveyed using
the narrowband filter NB2090 ($\lambda_{c}=2.095~\mu$m;
$\delta\lambda=0.019~\mu$m). Their data are extremely deep, with a
$5\sigma$ limiting magnitude of 24.6 AB corresponding to a line flux
limit of $6.8\times10^{-18}$ erg s$^{-1}$cm$^{-2}$, nearly an
order of magnitude deeper than our own data. The inset histogram in
their Figure~1 shows that the number of emitters they detected at
NB$<22$ is 9, under their EW cut of EW$_{o}>20~$\AA\ (corresponding to
the emission-line excess of 0.29). In their filter bandwidth our
selection criterion of EW$_{o}>50$~\AA\ corresponds to an emission-line
excess of 0.60 mag. We estimate that 5 of their emitters satisfy our
selection criterion of EW$_{o}>50$~\AA\ from their Figure~1. Considering
the difference in survey areas, the expected field count of the NB
emitters for our data is estimated to be $2.5\pm1$.  
This estimate is consistent with the results from another deep survey, carried out 
on a smaller field, by \citet{moorwood00}. Our count of 11 HAEs is therefore
$5.3^{+5.3}_{-2.6}$ times (or $3\sigma$) more than the field count,
which is based on the proper treatment of the small number Poisson
statistics by \citet{gehrels86}.

We note that the redshift evolution of the HAE number count suggested by \citet{geach10} is quite strong: extrapolating from redshifts below $\rm{z}=2.2$ to $\rm{z}=2.5$, the 
predicted number counts of HAEs with flux larger than $1~10^{-16}$\ erg\ s$^{-1}\ $cm$^{-2}$ decreases by $60\%$, from 8582 to $\sim$5380 $\rm{deg}^{-2}\rm{z}^{-1}$. Although the 
evolution in number density depends on the limiting flux, this 
correction for the difference in redshifts (from $\rm{z}=2.2$ to $2.5$) will 
also reduce our field count estimate based on Hayes et al.~(2010). Our 
estimate of the overdensity of a factor 5 should therefore be considered 
conservative.

As we show in Table~\ref{tbl2}, about half of our HAEs seem to have a
stellar mass of $\gtrsim10^{11}~{\rm M}_{\solar}$. Such massive objects at
  ${\rm z}\sim2.5$ are relatively rare in the field. For example, if we
  count such objects using the MODS catalog (\S3.5.3), we find only
  10 in their 103-arcmin$^{2}$ survey area at $2.3<{\rm z}<2.6$
  (Fig.\ref{fig10}). By scaling the value to our survey area of
  23.6 arcmin$^{2}$ and $2.47<{\rm z}<2.49$, we estimate the expected number
  in our FOV to be 0.15. The excess of such massive galaxies in the 
4C~23.56 field is clear. Recalling that these massive galaxies in the 4C~23.56 field
are H$\alpha$-selected, there could be more inactive massive 
galaxies missed by our search. We are planning to execute multi-wavelength 
observations of the 4C~23.56 field. The data will help identify such inactive massive 
galaxies, and will reveal the whole picture of the protocluster.

\subsection{Comparison with Other Protoclusters}

Although there are several known protoclusters at ${\rm z}>2$, 
there is only one other field that has been searched for HAEs.
\citet{kurk04a} searched for HAEs in their discovered protocluster surrounding
PKS~1138-262 at ${\rm z}=2.16$. Here we compare the number of 
detected HAEs.

According to their Figure~6 and Table~5, their selection of
EW$_{o}>50$~\AA~ and $\Sigma>3$ combined with a magnitude limit of
NB$_{\rm Vega}<20.2$ would match the selection we use here. There are
$\sim 10 $ HAEs meeting these criteria in the PKS~1138-262 field, roughly the
same number as the 4C~23.56 field.  Their HAE survey field is roughly half
of ours (but wider than our ``INGRID subfield'' where all our HAEs
lie).  However, if we use the distribution of their Ly$\alpha$
emitters as a proxy for the HAE distribution (their Fig.~10) there
would be at most an additional two HAEs.  This implies that the
richnesses of the 4C~23.56 and the PKS~1138-262 fields are likely
comparable.

There are differences in the properties of the HAEs in the PKS~1138-262
and the 4C~23.56 protoclusters, primarily the spatial distribution of the HAEs relative
to the radio galaxies.  In the PKS~1138-262 system the HAEs are rather
homogeneously distributed around the radio galaxy, with some evidence
for higher surface density closer to it.  In the 4C~23.56 field the
radio galaxy lies at the western edge of the HAE distribution.  This
does not appear to be a result of observational bias as described in
\S2.1.1 and \S3.2.

This may imply a different evolutionary future of the host galaxy of
each radio source.  For example, the PKS~1138-262 host may evolve into the
centrally-dominant cD-like galaxy relatively early while the 4C~23.56 host
may remain one of several dominant galaxies until they merge into a
single larger galaxy at the center of the potential. The host of
4C~23.56 is indeed $\sim1.5$ magnitudes less luminous in $K_{\rm s}$
than the PKS~1138 host ($K_{\rm s}=18.66$ after removing nuclear component
and emission-line contribution; \cite{pentericci97}).  Furthermore, the
$K_{\rm s}=18.66$ magnitude for the host of PKS~1138 is unusual; its apparent
magnitude is as bright as the brightest cluster galaxies at
${\rm z}=1.5$\ \citep{collins09}. If we use formula (\ref{eq6}) in \S3.5.3, the 
stellar mass of the 4C~23.56 and the PKS~1138 hosts are estimated to be
$4.7\times10^{11}$M$_{\solar}$ and $5.5\times10^{12}$M$_{\solar}$,
respectively. For comparison, \citet{seymour2007} estimated the stellar
mass of both host galaxies based on their {\it Spitzer} multi-band
photometry, finding $3.9\times10^{11}$M$_{\solar}$ and
$<1.3\times10^{12}$M$_{\solar}$, for 4C~23.56 and PKS~1138 respectively.

Another difference is in the number of HAEs with strong
(EW$_{o}>100$~\AA) emission. In the 4C~23.56 field, 10 of 11 HAEs
have rest-frame EW$_{o}>100$~\AA. In the PKS~1138 field the number of
objects that satisfy EW$_{o}>100$~\AA~ and NB$_{\rm Vega}$$<20.2$ is 5. As
the equivalent width relates to the H$\alpha$ luminosity, the excess
number of strong EW objects in the 4C~23.56 field implies more active star
formation occurring in that protocluster. We will address this quite
interesting topic more in the next section.

\vspace{1ex}

Recently, \citet{hayashi10} reported an excess of [O\emissiontype{II}] emitters
in the field of the distant X-ray cluster XMMXCS~J2215.9-1738 at ${\rm z}=1.46$.
The distribution of [O\emissiontype{II}] emitters are prevalent from the outskirt
to the core region of the X-ray cluster. They showed that the fraction of the [O\emissiontype{II}] emitters among all cluster members as well as the level of the star-formation are shown to be both nearly constant as a function of clustercentric radius (see also, \cite{tran10}).

Their Figs.5-7 indicate that the distribution of the emitters
well traces the overall distribution of all cluster members.  The
distribution of H$\alpha$ emitters in \citet{kurk04a} also closely
matches the photo-z-selected galaxy distribution as shown by their 
Fig.6 \citep{tanaka10}. 
We note, however, that the case of the cluster XMMXCS~J2215.9-1738 may not be the representative as a cluster at $\rm{z}\sim1.5$. \citet{hilton10} has shown that the system has a possible bimodal kinematic structure, with the position of the cluster BCG significantly offseted from the center of the thermal hot X-ray gas. And also, the NB HAE survey for another cluster XMMU~J2235.3-2557 at ${\rm z}\sim1.4$ by \citet{bauer10} has show that the HAE distribution just avoids the central $\sim200$ kpc of the cluster core (see also \cite{lidman08,strazzullo10}). Although we are still not sure about whether the HAE can trace the overall distribution of cluster galaxies from the core to the outskirts at $z>2$ universe, both studies could consider the supporting case that the distribution
of HAEs in the 4C~23.56 protocluster also traces the underlying
galaxy distribution.  Future analysis based on the broadband imaging
will enable us to address the issue.

\citet{hayashi10} contrast the activity of galaxies in the
cluster core region by comparing to similar work for a ${\rm z}=0.8$ cluster
by \citet{koyama10}. At even lower redshift, \citet{kodama04} studied
the HAE distribution for Cl~0024+1652 at ${\rm z}=0.4$, showing quite strong
suppression of activity in the cluster core region. It is quite
interesting to connect the results by these works from ${\rm z}=0.4$\ \citep{kodama04},
${\rm z}=0.8$\ \citep{koyama10}, ${\rm z}=1.4-1.5$\ \citep{hayashi10, bauer10},
${\rm z}=2.1-2.2$\ \citep{kurk04a, gobat10}, to our ${\rm z}=2.5$ studies as a snapshot of the cluster evolution in each epoch \citep{tadaki10}. The implied rise of star-formation activity in the cluster core with redshift and its variety suggests that, while it may indicate that we are gradually approaching the era of galaxy formation for members in the cluster core \citep{kodama01}, it is important to evaluate the physical state (richness, dynamics, mass etc.) for each protoclusters in detail to draw a general pictures of early history of galaxy evolution in cluster core.  The detailed studies of larger samples are clearly needed to corroborate this apparent evolutionary trend.

\subsection{Star-Formation Activity of HAEs}

In \S3.5 we calculate the SFR and the stellar mass of our HAE
sample.
Although we admit that the estimate of these values is
rather crude, here we try to characterize the star-formation of the
HAEs in 4C~23.56 field by comparing them to the other studies.
As is discussed in \S3.5, the main uncertainty for H$\alpha$-based
SFR is in the estimate of the dust extinction, as indicated by the MIPS 
detection. The unknown fraction of the AGN contribution to our HAE sample 
(it can be non-negligible: see \cite{lehmer09, digby10, lemaux10}) is also an 
issue we need to address by the future observation.
%
%

In Figure~\ref{fig9}a, we plot the estimated star-formation rate of
our HAEs against their stellar mass. The minimum SFR in our sample is
$\sim 15~{\rm M}_{\solar}~{\rm yr}^{-1}$, with a median value of 45 (182 for
extinction corrected, 746 for MIPS) ${\rm M}_{\solar}~{\rm yr}^{-1}$. In the
figure, the brightest object with the highest SFR, is 4C~23.56 (note
that the mass estimate for this object is probably overestimated, not
only due to the inclusion of AGN but also due to the effect of the
poor sky subtraction: see \S2.1). The line on the figure is the
average mass-SFR relation for ${\rm z}\sim2$ BzK-selected galaxies with
MIPS detections by \citet{daddi07}. 
The SFRs based on both indicators show larger values than expected from the 
${\rm z}\sim2$ mass-SFR relation by \citet{daddi07}. 
Especially, the MIPS-based SFR shows large excess. This 
trend seems real, because our MIPS-based SFR is derived from the same equation
as used by \citet{daddi07}. This may indicate that some galaxies in the protocluster 4C 23.56 have 
higher star-formation activity than the field counterparts with similar mass. 

The observed large excess of the $24\mu m$-based SFR relative to $H\alpha$-based one 
may indicate that the star-formation activity in our HAE sample is largely (or 
selectively) hidden by dust. At lower redshifts such hidden star-formation activity
revealed by infrared observation is often reported (e.g., \cite{valdes05, geach06, garcia09, garn10}). On the other hand, the recent calibration of the MIPS $24\mu m$-based 
SFR for star-forming galaxies at ${\rm z}\sim2$ by the {\it Herschel} PACS observation
has revealed that the SFRs from $24\mu m$ alone could be overestimated by a factor 
of $\sim 2-7.5$ \citep{nordon10, rodighiero10}. Our observed
trend can be due to the large systematics in the estimate of the total infrared luminosity
from the rest-frame $8\mu m$ photometry. For now we do not know whether the HAEs in the 
4C 23.56 field harbours the star-formation significantly hidden by dust. The direct 
far-infrared observation by ALMA will address the issue.

In Fig.\ref{fig9}b we show the L(H$\alpha$)-based specific star-formation rate 
(SSFR) as a function of stellar mass.  For comparison, we plot the SSFR
values for field H$\alpha$ emitters from \citet{sins09} and
\citet{onodera10}. The \citet{sins09} points include two types of the
extinction correction, which is shown as connected points with vertical
lines in the figure, while \citet{onodera10} used the correction
(described above) which could be considered the middle between the two
extreme cases from \citet{sins09}.

The measured relation between SSFR and mass deviates clearly from that 
observed in similar studies of low redshift clusters.  For example, 
\citet{patel09} measured the SFR and SSFR of a
rich cluster at ${\rm z}\sim0.8$ from MIPS-based photometry, and confirmed
that the SSFR of massive (log~${\rm M}/{\rm M}_{\solar}>10.8$) galaxies in high
density regions is as low as for field objects with similar mass in
local universe. 
Also at ${\rm z}>1$, the SSFR estimate of the member galaxies in the irregular 
cluster CL 0332-2742 at
${\rm z}\sim1.6$ by \citet{kurk09} show that significant member galaxies have
much lower SSFR than field counterparts at similar redshifts, despite that the
system is quite irregular and far from a virialized structure.
Recently, \citet{tanaka10} studied galaxies in the
protocluster around PKS~1138 at ${\rm z}=2.16$ and showed a similar trend of the 
suppressed star-formation activity in the cluster compared to the field
environment (see also, \cite{zirm08}). These studies may indicate that
the star-formation activity in cluster environment is suppressed up to
${\rm z}\sim2$.

To avoid the uncertainties introduced by estimating the SFR and the
stellar mass, we also compare the star-formation rates directly using
the observed values.  In Figure~\ref{fig11} we show the observed
$K$-band magnitude versus the H$\alpha$ luminosity ($L($H$\alpha$)) of
the same dataset as shown in Fig.\ref{fig9}b, but limited to objects
at ${\rm z}>2$. Here the data from \citet{kurk04a} is also added for
comparison. Within the uncertainties, all HAEs in the 4C~23.56 field are clearly
as luminous as the brightest H$\alpha$ sample in the general field
results, again suggesting an excess of active star formation in the 4C~23.56 field.
%
%

Perhaps more interestingly, in Fig.\ref{fig11} we compare the HAEs in the 4C~23.56
field with the PKS~1138 field data. The majority of HAEs in the PKS~1138 protocluster 
have lower
H$\alpha$ luminosities than both field objects and the HAEs in
4C~23.56 despite being within a similar $K$ magnitude range. It
indicates that the HAEs in the PKS~1138 field show suppressed star
formation. This is actually the same conclusion reached by
\citet{tanaka10}, but their work is based on broadband SED fitting,
i.e., from an independent measure.

The cosmic time between the two protoclusters is only 0.4~Gyr.  This is
actually long enough to discern changes of the star-formation activity
if their star-formation timescale is short.
\citet{tanaka10} suggest that the star-formation timescale for
galaxies in the PKS~1138 field is much shorter than for galaxies with
similar mass in the GOODS fields. Their Figure~13 demonstrates the
rapid decline of the SFR in the protocluster compared to that of
GOODS, and shows that the time difference of 0.4 Gyr is long enough to
observe the decline. For example, if we suppose that these
star-forming galaxies in protoclusters are formed at ${\rm z}\sim3$, the
increase of the star-formation rate between ${\rm z}=2.16$ to 2.5 (0.4 Gyr)
is close to an order of magnitude.

This argument could of course be too simplified, as it assumes that
any galaxies in the two clusters have almost the same formation period
around ${\rm z}\sim3$. The actual formation of massive galaxies in the
cluster environment could be much more complicated, and it could be
coupled with the fundamental cluster properties (mass, merger history,
dynamical state, and so on). As we already discussed, the difference
of the star-formation activity between the two protoclusters might be
related to the different inherent cluster characteristics. The
PKS~1138 host is exceptionally massive and already lies near the
apparent center of the galaxy distribution \citep{kurk04a,
  seymour2007}, while the 4C~23.56 host lies at the edge of the HAE
(and presumably overall galaxy) distribution (\S3.2 \& \S4.2). The
soft X-ray emission from the hot intracluster gas is (marginally)
detected in PKS~1138 \citep{carilli02}, while for 4C~23.56 it is
not \citep{johnson07}. The future of the PKS~1138 protocluster may
well be an X-ray bright, well evolved rich cluster with a massive cD
galaxy lying in the center, while that of 4C~23.56 might be an X-ray
faint cluster with a few dominant giant ellipticals near the center of
the potential, such as the Coma or Virgo clusters.

\section{Summary}

We have discovered a significant excess of near-infrared selected
emission-line objects around the radio galaxy 4C~23.56 at
${\rm z}=2.486$. We detect 11 robust emission-line galaxies with a line flux
greater than $7.5 \times 10^{-17} {\rm erg}\ {\rm s}^{-1}\ {\rm
  cm}^{-2}$ in our MOIRCS imaging observation of the $4^{\prime}
\times 7^{\prime}$ field around the radio galaxy. We show that this is
a significantly higher (5 times more) detection rate compared to other
near-infrared emission-line surveys in non-radio-galaxy fields. Three
of the emission-line galaxies are found to have a spectroscopic
redshift of ${\rm z}=2.49$.
We therefore conclude there is a protocluster physically associated with
4C~23.56.

The distribution of HAEs on the sky is confined to a relatively small
area ($\sim1.2$Mpc-radius region) near the radio galaxy, with 4C~23.56
placed at the western edge. We show that this is not due to
observational bias, and conclude that the protocluster is
physically associated with the radio galaxy. The offset position of the
radio galaxy is interesting, because the hosts of powerful radio
galaxies are usually very massive and often assumed to be the dominant
object (e.g., cD galaxy) in the center of the cluster. We also find
that there is an interesting compact group of four HAEs, all the
members of which show offset H$\alpha$ emission relative to the
$K_{s}$-band light position. This may indicate strong star-formation
on the edge of each galaxy which is occurring due to strong
interactions with the group. A more exotic interpretation might be that
there is ram-pressure stripping of the galactic-scale
H$\alpha$-emitting gas by the hot intracluster gas.  Further
follow-up is necessary to understand the phenomenon.

Our analysis of deep {\it Spitzer} MIPS $24~\mu{\rm m}$ data
for 4C~23.56 shows that all but two of the HAEs have 24$~\mu{\rm m}$
counterparts. The inferred star-formation rate (SFR) based on the MIPS
24$~\mu{\rm m}$ photometry is quite high (median of $>700~{\rm M}_{\solar}~{\rm yr}^{-1}$), indicating that intense starburst activity is going on in each HAE. The
MIPS-based SFR is on average 3.5 times higher than the value inferred
from the observed H$\alpha$ luminosity. The strong effect of ``selective''
dust extinction is evident for these galaxies, implying the metal-rich
and evolved populations for HAEs, as suggested by \citet{kurk04a}.

The analysis of the spatial distribution of the MIPS
sources has also revealed that there exists a strong excess around
4C~23.56. Notably, the distribution of sources that have
$24~\mu{\rm m}$ fluxes as faint as the HAEs ($90-275~\mu {\rm Jy}$) shows the same
offset distribution relative to the radio galaxy, and the position
of the density peak nicely coincides with the HAE
distribution. This is another indication that the strong star-formation
activity is going on in the protocluster.

The comparison of the star-formation rates of HAEs in the protocluster
4C~23.56 at ${\rm z}=2.48$ to those in another protocluster PKS~1138-262
at ${\rm z}=2.16$ and to those in the general field at ${\rm z}\sim2$ shows that:
(1) HAEs in the 4C~23.56 field are as active as those in the general field,
and that (2) HAEs in the protocluster PKS~1138-262 show a clear {\it deficit} 
of objects with high H$\alpha$ luminosities.
The clear difference in star forming activities between the two protoclusters
at similar redshifts may indicate that we are stepping into the epoch of major
star formation in massive cluster galaxies.
Given the fact that there is a small but significant time difference
of 0.4~Gyr between the two redshifts, such variation in star formation activity
may be interpreted as a time evolution where star formation activities
to form massive galaxies can decay rather sharply during this epoch
($2<{\rm z}<3$) in proto-clusters (e.g., \cite{kodama07, tanaka10}).
Alternatively, we may be just seeing the intrinsic scatter of cluster
properties at this high redshift where the onset of cluster formation 
and the subsequent evolution depend on the mass of the systems and/or 
structures in the surrounding environment.
Such dependence is actually inferred in lower-redshift clusters at ${\rm z}<1$ 
(e.g., \cite{urquhart10}).
However, we do not know yet the global properties of these protoclusters,
such as total number of member galaxies, cluster masses, and the dynamical
states, nor do we yet have a statistical sample of protoclusters at
this interesting redshift regime to disentangle these two effects.
We are desperate to obtain such information by extensive near-infrared 
spectroscopy and to increase the sample of proto-clusters at ${\rm z}>2$.

\bigskip

We thank the anonymous referee for comments that has lead to the improvement of the paper. We thank the MOIRCS Deep Survey Team for providing us with the MODS catalog. Also, we would like to thank Dr.~Sune Toft and Dr.~Masa Tanaka, and many others for useful comments and suggestions. We also thank Dr. Ryuji Suzuki for giving simulation data on the shift of the central wavelength of NB filters. This work was supported by the JSPS Institutional Program for Young Researcher Overseas Visits. YT and TK thank for support via Grant-in-Aid for Scientific Research (KAKENHI: Nos.\ 19340046 and 21340045, respectively) by the Ministry of Education, Culture, Sports, Science and Technology in Japan. JK thanks the DFG for support via German-Israeli Project Cooperation grant STE1869/1-1.GE625/15-1.  IT would like to thank the all ESO staff for the hospitality during the stay in ESO Garching. IT would also like to thank all the staffs and members of the DARK Cosmology Center, MSSL University College London, and Durham University for their warm hospitality during my stay related to this study. 
This work is based on data collected at Subaru Telescope, which is operated by the National Astronomical Observatory of Japan. This is alos based in part on observations made with the {\it Spitzer} Space Telescope which is operated by the Jet Propulsion Laboratory, California Institute of Technology under a contract with NASA, as well as the data collected by the 4.2m William Herschel Telescope which is operated on the island of La Palma by the Isaac Newton Group in the Spanish Observatorio del Roque de los Muchachos of the Instituto de Astrof{\'i}sica de Canarias.

\newpage

\appendix
\section{The $BJK$ Color Selection for Rejecting Objects at ${\rm z}<1$}
In order to reject the possibility that our single line detection by NB filter or the spectra is mainly from the low-redshift Paschen series interlopers, we introduce the broadband color selection by the $B-J$ versus $J-K_{\rm s}$ two-color diagram. Here we show its effectiveness using the large photometric/spectroscopic catalog for the GOODS-North region. 
 
\begin{figure*}
\begin{center}
\FigureFile(75mm,75mm){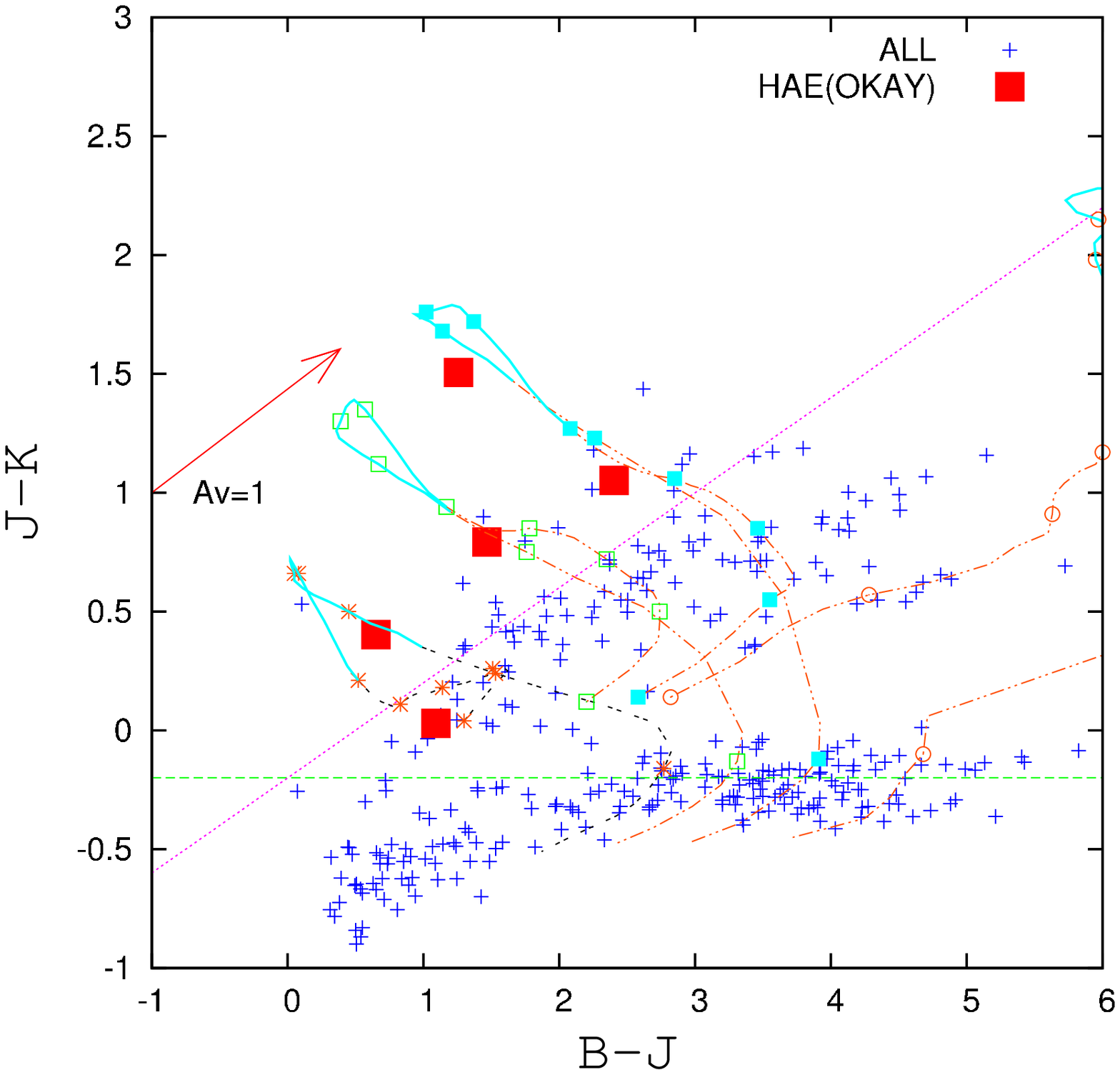}
\FigureFile(85mm,85mm){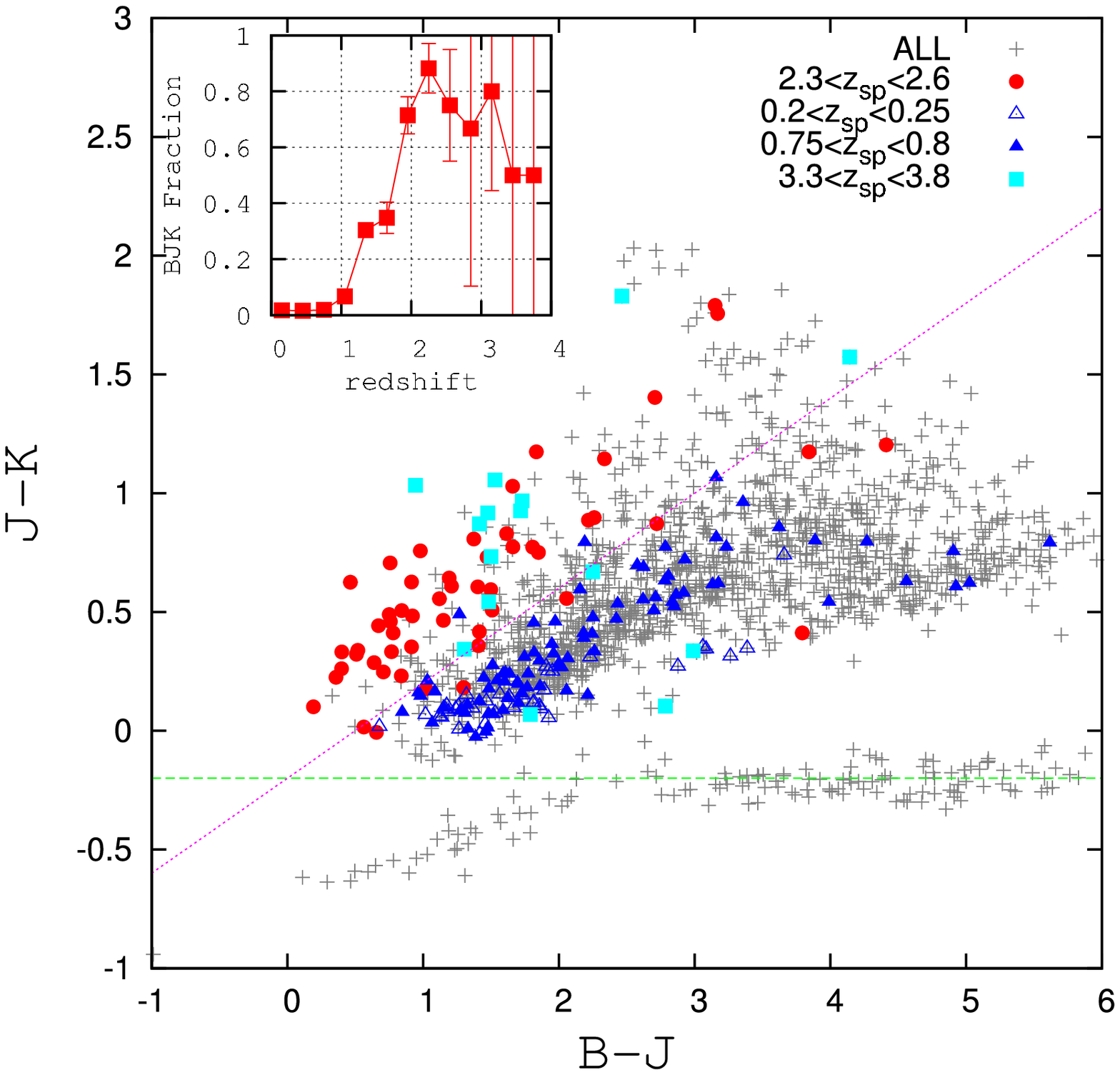}
\end{center}
\vspace{3ex}
\caption{{\bf (a: left)} The $B-J$ versus $J-K_{\rm s}$ two-color diagram for objects in the INGRID subfield. Five red boxes with the label ``HAE(OKAY)'' indicate the HAE candidates for which we can measure colors. Small plus symbols indicate other NB-selected objects in the area. The overplotted model tracks from blue to red indicate the ``pure disk'' model with continuous SFR (bluest), ``50\% disk + 50\% bulge'', ``20\% disk + 80\% bulge'', ``pure bulge'' (reddest, and mostly out of the figure) models, respectively. The cyan-colored region on each model track indicates the range $2.0<{\rm z}<3.8$, and the farthest point in each cyan-colored track is around ${\rm z}=2.5$ to 3. See text for more details.
{\bf (b: right)} The same figure for galaxies in GOODS-North field (see Appendix for details). The data is from the MOIRCS Deep Survey (MODS) Catalog with $K_{\rm s} < 23$ mag \citep{kajisawa09, kajisawa10}. In the figure, red filled circles indicate objects with spectroscopic redshifts of $2.3 < {\rm z} < 2.6$, similar redshifts to our H$\alpha$ emitters at ${\rm z}\sim2.5$. Blue triangles indicate objects at $0.2<{\rm z}<0.25$ (open) as well as $0.75<{\rm z}< 0.8$ (filled), each of which corresponds to the redshift range for our foreground Paschen series (Pa$\alpha$ and Pa$\beta$, respectively) interlopers, and the filled boxes (skyblue) indicate objects at $3.3 < {\rm z} < 3.8$ to which the background [O~{\sc iii}] emitter contaminant for our data would belong. Inset histogram shows the fraction of objects with known spectroscopic redshift that satisfy the $BJK$ criteria we used here as the object selection at ${\rm z}>2$. Note that there is a very limited number of spectroscopically confirmed galaxies at ${\rm z}>2.7$, as indicated by the large N$^{1/2}$ errorbars.\label{fig3}}
\end{figure*}

\begin{figure*}
\begin{center}
\FigureFile(160mm,100mm){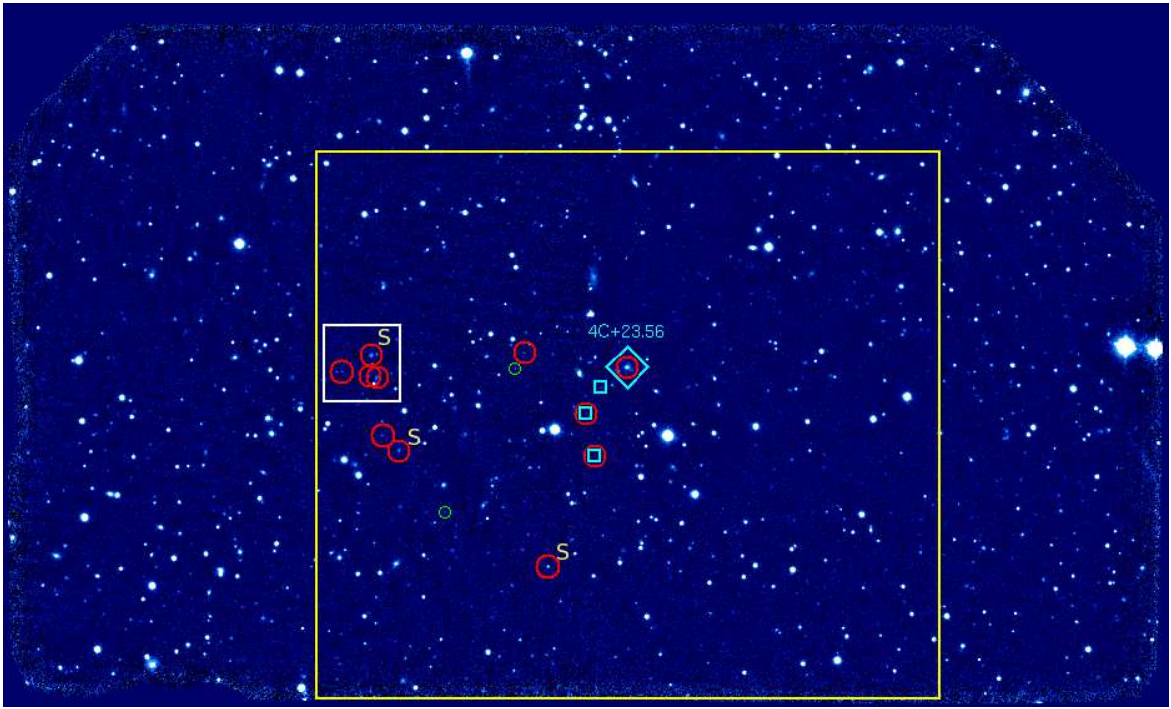}
\end{center}
\vspace{0ex}
\caption{The distribution of the HAEs on the NB2288 image. The FOV covers $7\arcmin \times 4\arcmin$ on the sky, with north and east to the top and left, respectively. The red open circles indicate the candidate HAEs by our selection. The three small open boxes indicate the Distant Red Galaxies from \citet{kajisawa06a} that show emission-line excess ($K_{\rm s}-$NB$>1$) as seen in Fig.{\ref{fig12}} (also see Appendix). Two of them are also identified as HAEs. The remaining one is the only NB-excess object among the ERO sample by KC97. Two small open circles (green) indicate the candidate emission-line objects that satisfy the criteria of $\Sigma >2$ but are fainter than NB=22.0. Three red open circle with the letter~"S" indicate the emitters with the MOIRCS spectra (\S3.1.1).  4C~23.56 is shown as a large diamond. The area enclosed by the white box indicates the same area shown in Fig.\ref{fig7}. The yellow rectangular region indicates the INGRID subfield (see \S3.1).\label{fig6}}

\end{figure*}

\begin{figure*}
\begin{center}
\FigureFile(160mm,100mm){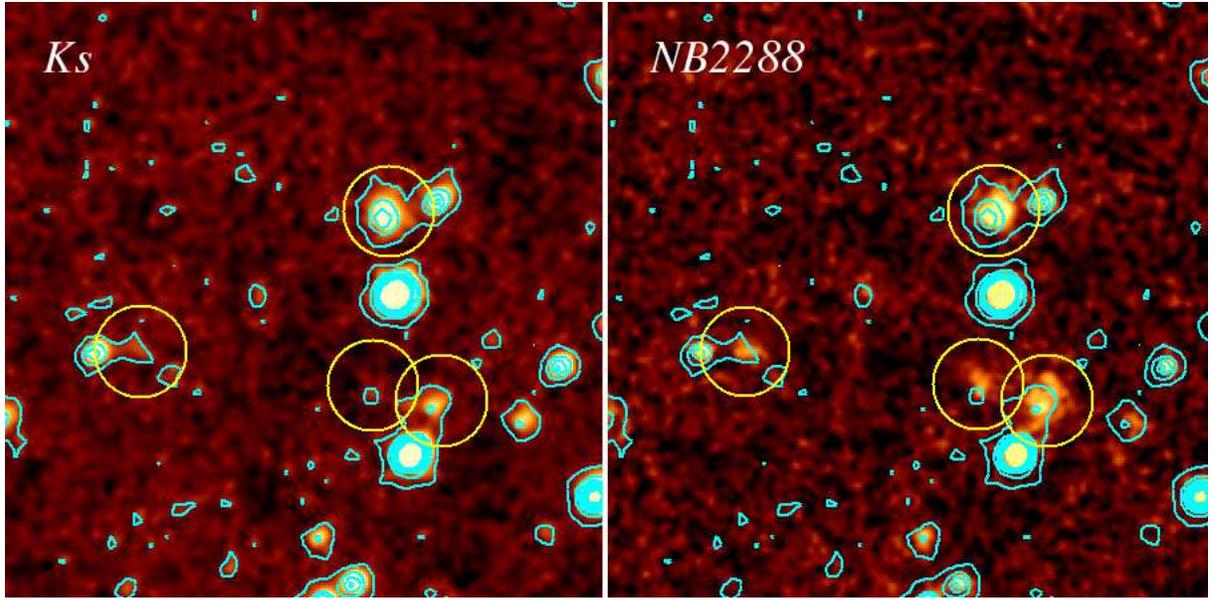}
\end{center}
\caption{Zoom-up views of the ``Eastern clump'' region in both the NB2288 (right) and the $K_{\rm s}$ (left) images. The objects at the center of each yellow circle represent the HAEs. The $1\sigma, 4\sigma, 7\sigma, 10\sigma, \& 13\sigma$ contours based on the rms noise of the $K_{\rm s}$ image before smoothing is applied to the Gaussian-smoothed $K_{\rm s}$ images, and is also overlaid on the NB2288 image. A clear extended or offsetted light distribution of HAEs in the NB2288 image is seen for (at least) the three objects on the right. Interestingly, the directions of the offsetted light distributions for these three objects are roughly the same (to the North-West). The size of the area is about $26\arcsec\times26\arcsec$ ($\sim0.5\times0.5$ Mpc$^{2}$ comoving).\label{fig7}}

\end{figure*}

\begin{figure*}
\begin{center}
\FigureFile(120mm,100mm){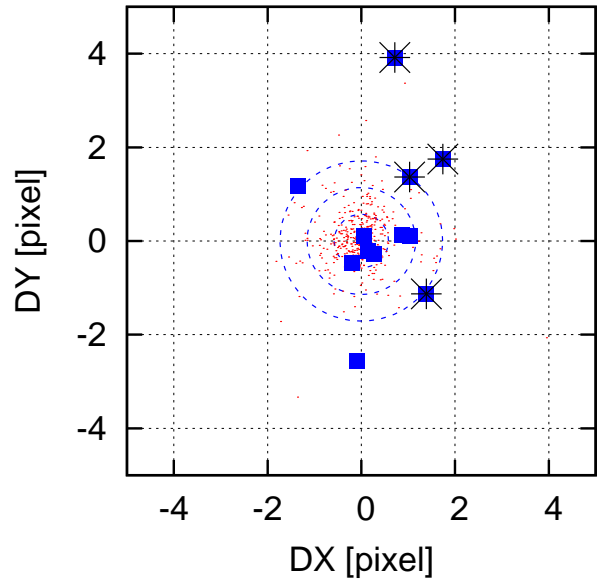}
\end{center}
\caption{The difference of the object position between the $K_{\rm s}$ and the NB2288 images. Objects that have similar magnitudes to our HAEs are used (dots). The 1-, 2-, and 3-$\sigma$ scatter of the distribution are shown as dashed circles in the figure from the inner to outer, respectively. The result for the HAEs is shown by the filled squares, with the 4 HAEs in eastern clump in Fig.\ref{fig7} overlaid by asterisks. The offset is shown in X and Y pixel coordinates, where the MOIRCS pixel scale is $\timeform{0.117"}/$pixel (corresponding physical scale is 0.96 kpc/pixel at ${\rm z}=2.48$). \label{fig8}}
\end{figure*}

\begin{figure*}
\begin{center}
\FigureFile(150mm,100mm){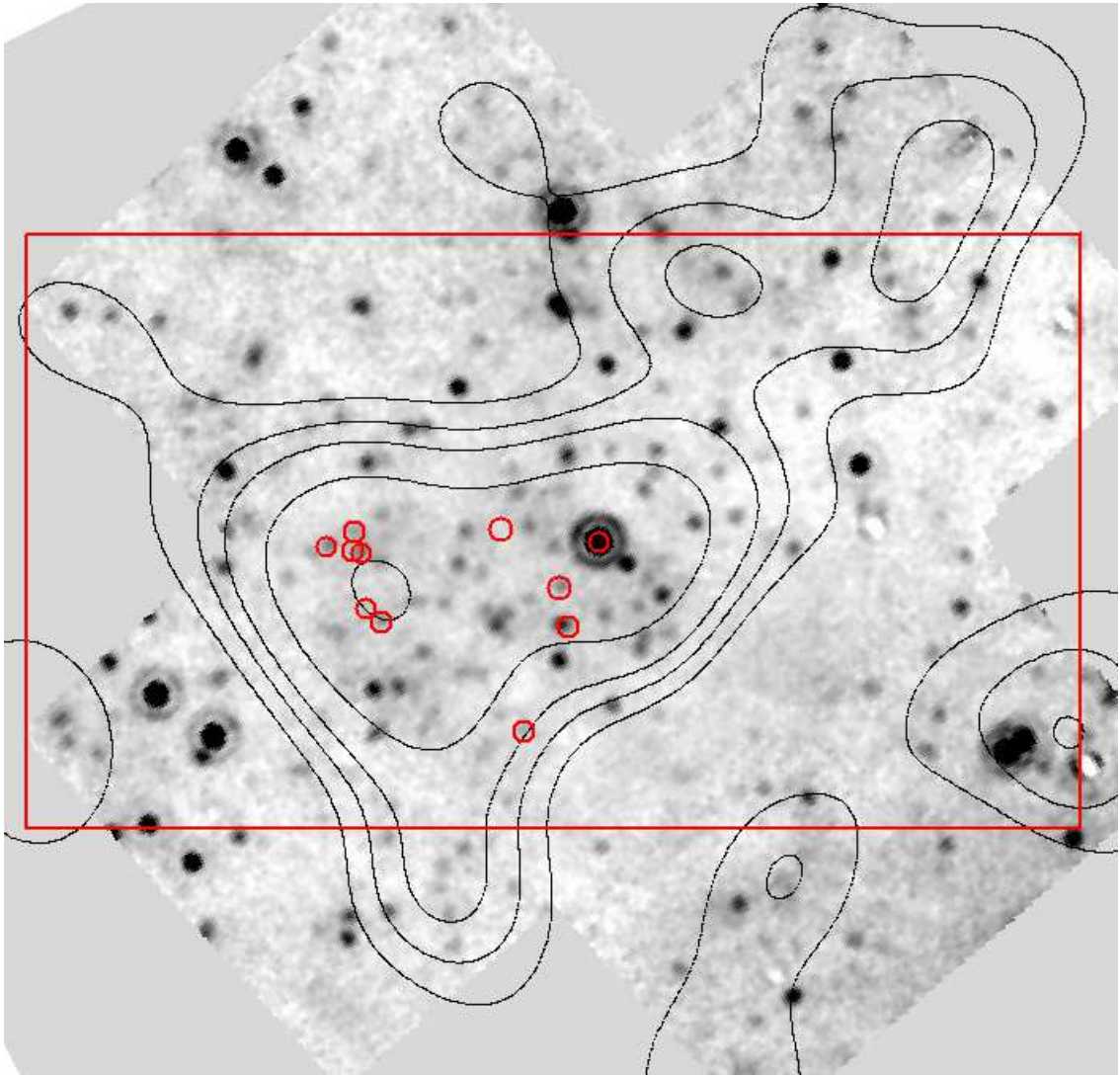}
\end{center}
\caption{The density contour of the faint {\it Spitzer} MIPS 24$\mu$m sources is overlaid on the MIPS 24$\mu$m image. The area with an exposure coverage less than 1/3 of the total is not shown here. The brightest source in the center is 4C~23.56. The red rectangular region indicates the MOIRCS field of view (North is up: see Fig.~\ref{fig6}), and the red open circles indicate the HAEs. The distribution of the faint MIPS sources ($90 < F_{\rm 24\mu{\rm m}}~[\mu{\rm Jy}] < 275$) is smoothed with a Gaussian kernel with $\sigma=30\arcsec$, and shown as contours. The contour interval is $(2i+1)$-times ($i=0,1,3,6,11$) the square root of the average MIPS source density (from the ``outer region'': see text) above average.\label{fig13}}
\end{figure*}

\begin{figure*}
\begin{center}
\includegraphics[scale=0.6, angle=270]{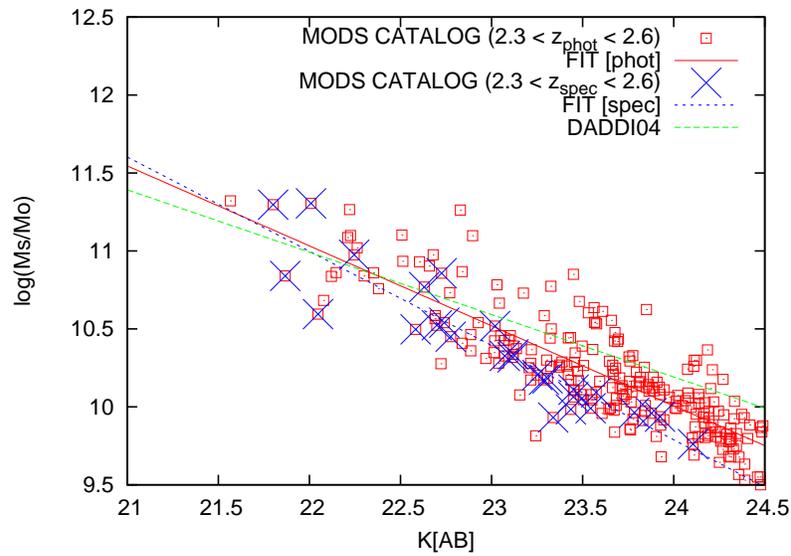}
\end{center}
\caption{The $K_{\rm s}$ versus stellar mass relation for objects at $2.3<{\rm z}<2.6$ from the MOIRCS Deep Survey (MODS) catalog used in \citet{kajisawa09}. The catalog includes the compilation of the spectroscopic redshift data published, and are shown as crosses in the figure. The stellar mass is based on the SED-fitting by the 12-band photometry from UV to infrared, with the SED model from Bruzual \& Charlot (2003). Details of the model fitting procedure can be found in \citet{kajisawa09}.\label{fig10}}
\end{figure*}

\begin{figure*}
\begin{center}
\FigureFile(80mm,100mm){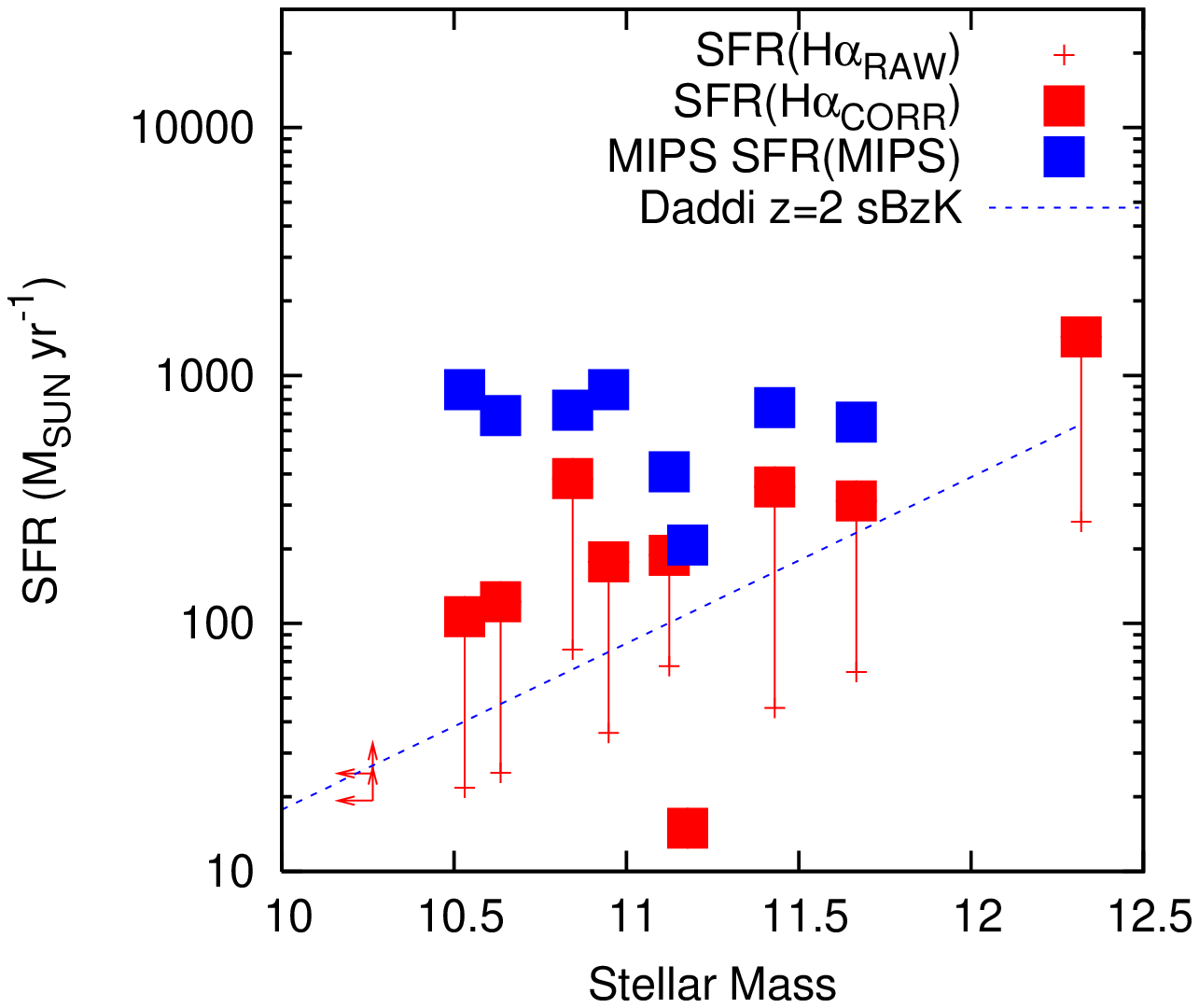}
\FigureFile(80mm,100mm){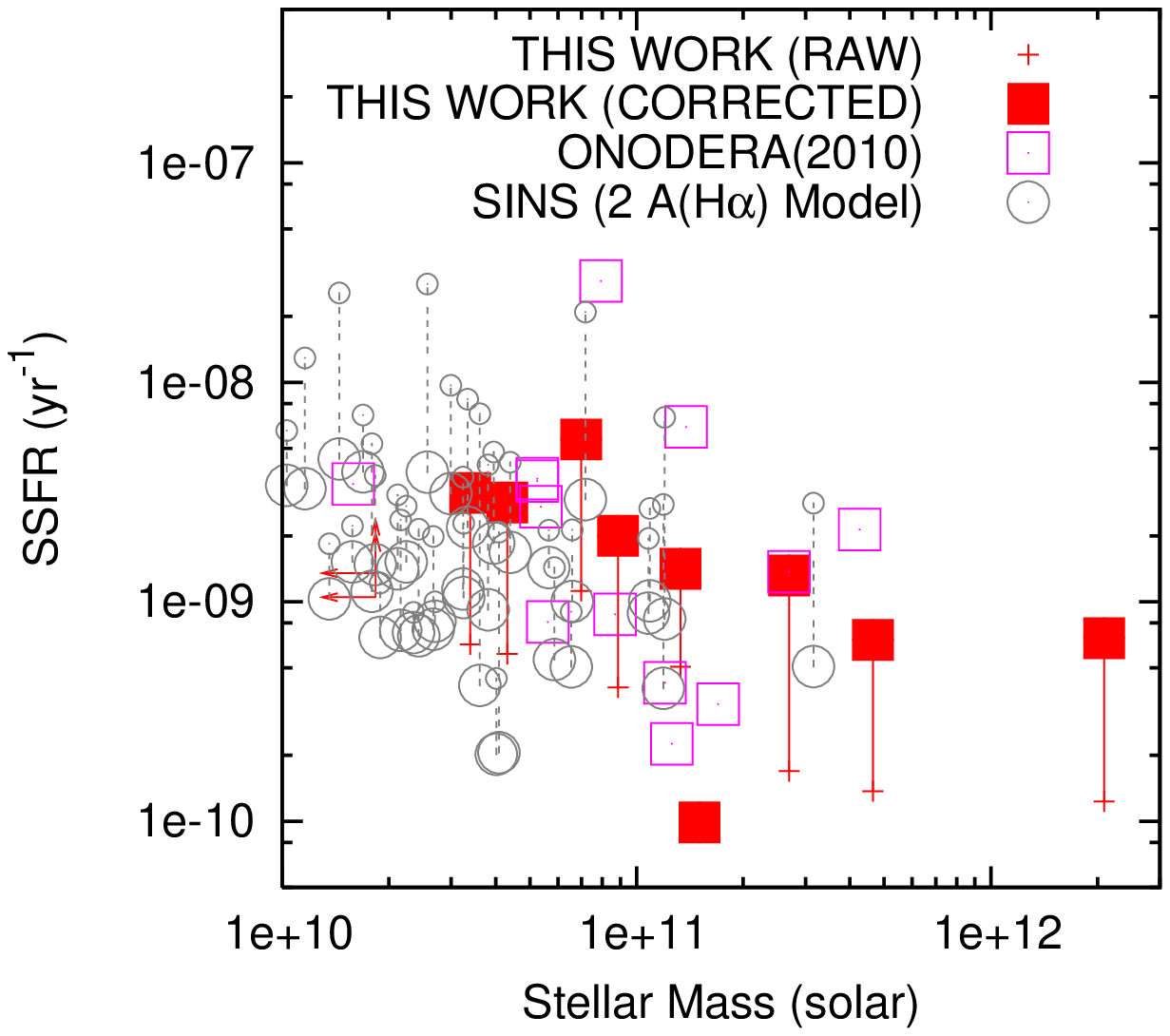}
\end{center}

\caption{{\bf (a: left)} The observed star-formation rate (SFR) of the HAE sample as a function of stellar mass. The plus marks indicate the H$\alpha$-based SFR \textit{before} the extinction correction, while red filled box connected to each plus mark is its extinction-corrected SFR values (see text). The blue filled boxes are the SFR based on the {\it Spitzer} MIPS $24~\mu{\rm m}$ photometry. The dotted line (blue) indicates the mass-SFR relation at ${\rm z}\sim2$ from \citet{daddi07}. The two arrows indicate the objects with the upper-(lower-)limit for their mass (SFR before extinction correction) due to their faintness. The most massive object is 4C 23.56. Its MIPS-based SFR is not shown (due to an AGN contribution and the model limit).
{\bf (b: right)} The comparison of the H$\alpha$-based specific SFR (SSFR) of our HAE sample and of some field H$\alpha$-emitting objects as a function of stellar mass. The marks with dust extinction correction (red filled box) and raw values (plus) are connected by vertical lines. The field samples are from \citet{sins09} and \citet{onodera10}. Note that \citet{sins09} shows the results using two types of the extinction correction; thus they are shown as connected vertical lines in the figure.
Our SSFR estimate should be treated as the crude values due to the large uncertainty in both the extinction correction and the mass estimate. Note that 4C~23.56 is the point at $M>10^{12}~{\rm M}_{\solar}$, and should be ignored because of strong AGN contamination in both the $K$-band and the H$\alpha$ emission.\label{fig9}}

\end{figure*}

\begin{figure*}
\begin{center}
\FigureFile(120mm,100mm){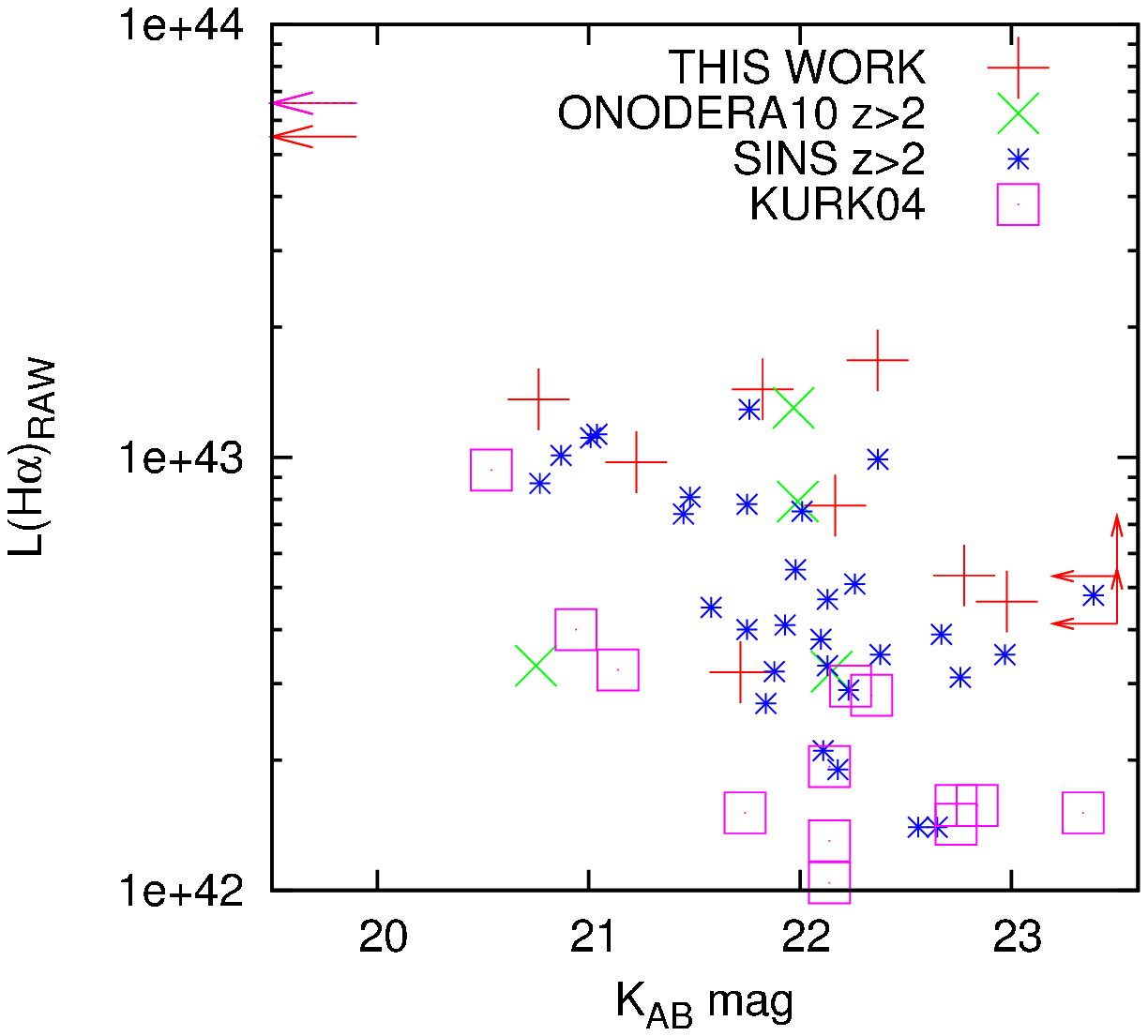}
\end{center}
\caption{Comparison of the $K_{\rm s}$ magnitude and $L($H$\alpha)$ for H$\alpha$ emitters from two field surveys from \citet{sins09} and \citet{onodera10} (asterisk and cross), PKS~1138-262 field from \citet{kurk04a} (box), and our study (plus). Only objects at ${\rm z}>2$ are shown for field survey data. For the PKS~1138 field, we plot the data with the significance level over 3 and EW$_{o} > 50$\AA\ for better match to our selection criteria. Two arrows at brighter than $K=20$ represents the PKS~1138 and 4C~23.56 radio galaxies which are out of the figure due to their bright magnitudes.\label{fig11}}

\end{figure*}

We make use of the MOIRCS Deep Survey Catalog which contains 13-band photometric data as well as the extensive compilation of spectroscopic redshift data (Kajisawa private com.; \cite{kajisawa10}). In Figure~\ref{fig3}a, we plot the color of all galaxies with $K_{\rm s}<23$ in GOODS-N (gray crosses). The galaxies at $0.2<{\rm z}<0.25$ and $0.75<{\rm z}<0.8$ (spectroscopically measured), each of which includes the Pa$\alpha$ and Pa$\beta$ interlopers, are shown by blue filled triangles, while the objects at $2.3<{\rm z}<2.6$, which include the 4C~23.56 redshift, are shown by red filled circles. There is a clear line between the two samples. Here we define it as,
\begin{equation}
(J-K_{\rm s}) = 0.4~(B-J)-0.2,
\end{equation}
which is shown as the magenta dotted line in the figure. The insetted diagram in Fig.\ref{fig3}b shows the fraction of galaxies whose $J-K_{\rm s}$ color is redder than the line defied here (``$BJK$'' galaxies) as a function of redshifts (spectroscopic redshift sample only). Almost no objects at ${\rm z}<1.2$ can be $BJK$ galaxies, while more than $80\%$ of objects at ${\rm z}\sim2.5$ satisfy the criterion. 

We also calculate the behavior of galaxy colors in the $BJK$ plane using the population synthesis SED models by \citet{ka97} overplotted in Fig.\ref{fig3}a. In the figure, each model track indicates, from bluer to redder, the ``pure disk'' model with continuous SFR (bluest), ``50\% disk + 50\% bulge'', ``20\% disk + 80\% bulge'', and ``pure bulge'' model with the formation redshift of 5 (reddest, and mostly out of the figure). In each model, symbols are put from ${\rm z}=0.1$ to 4, with the interval of 0.5 from ${\rm z}=0.5$ (each line starts from ${\rm z}=0.1$ with a symbol, and ends at 4.5 without symbol). A part of each model track has the region with cyan-colored solid line, which indicates the region at $2<{\rm z}<3.8$. The extinction vector for A$_{V}$=1.0 based on \citet{calzetti01} is also plotted.

The behavior of the model tracks also confirms that the our proposed dividing line for $BJK$ galaxies works rejecting objects at ${\rm z}<1$. The model tracks at $2.3 < {\rm z} < 2.6$ (the same redshift range as red filled circles in Fig.\ref{fig3}b) tend to go a bit further from the $BJK$ line than the actual distribution of objects at the same redshift range in Fig.\ref{fig3}b. This may be explained, though, by a difference in the metallicity between the modeled and observed galaxies. More advanced modeling is beyond the scope of this paper, however.

\section{Objects Identified in Previous Studies}
As is described in \S1, the 4C~23.56 field was also observed previously by KC97, \citet{kajisawa06b}, and \citet{stockton04}. Here we briefly comments on the objects detected in each study.

\subsection{Objects in KC97}
KC97 studied the 1.25 arcmin$^{2}$ region around the radio galaxy using deep $U'VIK$ imaging data from 2.2- \& 3.6-m telescopes. They also searched for the Ly$\alpha$ emitters using the narrowband filter at 4229\AA\ on the UH 2.2-m telescope. They cataloged 9 possible Ly$\alpha$-emitting companions and 15 EROs selected by $I-K_{\rm s}$ colors around the radio galaxy.

We first looked for the counterparts of their Ly$\alpha$ emitters in our $K_{\rm s}$ image. However, we can only identify one object (ID\#110 in their catalog). The $K_{\rm s}-NB$ color of the object is 0.14, so no clear sign of NB excess is seen in Figure~\ref{fig12} (red filled box).

Next we identify their EROs. Of their 15 objects, 6 objects are not seen at all in our $K_{\rm s}$ nor NB images, and 5 objects are quite faint and subsequently not in our SExtrctor catalog. Note that their data is extremely deep, with a $3\sigma$ limiting magnitude of $K=24.3$ mag, while our data only reaches 23.5 mag. Among the 4 remaining objects, only one object (their ID\#70) shows a sign of emission-line excess (green open circles in Fig.\ref{fig12}). However, the object is just below our HAE selection criteria (NB$<22$mag and $\Sigma > 2$) and therefore not present among our HAE candidates.

We comment on the failure of the HAE counterpart detection for their Ly$\alpha$ emitters in detail. We can calculate the expected H$\alpha$ flux from their cataloged equivalent width, assuming the Ly$\alpha$/H$\alpha$ ratio of 8.7 for the Case B recombination \citep{osterbrock89}. We find that the expected H$\alpha$ flux are mostly several $\times 10^{-17}$~erg\ s$^{-1}$\ cm$^{-2}$, close to our detection limit. However, this should be seen as a lower limit, as the Ly$\alpha$/H$\alpha$ can be much smaller than 8.7 due to the more effective dust absorption by the resonant scattering nature of the Ly$\alpha$ line \citep{kurk04a, hayes10b}. Thus at least some, if not all, of their 9 LAEs could be detected. We note that the cataloged $K$-band magnitudes of Ly$\alpha$ emitters are all quite faint (5 are upper limits with $K_{\rm AB}>24.35$, and only two are within our magnitude limit of $K_{\rm AB}>23.5$). The non-detection by our $K$ image could be reasonable. However they are also totally absent even in our NB2288 image.

\citet{kurk04a} also did not detect H$\alpha$ emission for their sample of Ly$\alpha$ emitters in the PKS~1138 protocluster. For two of their three brightest Ly$\alpha$ emitters, for which H$\alpha$ emission could be detectable, faint emission was seen, but not strong enough to pass the HAE selection criteria. The range of the upper limit of the Ly$\alpha/$H$\alpha$ ratio for their H$\alpha$ emitters are $0.03 - 1.06$, significantly lower than the Case-B assumption (see also, \cite{hayes10b}).

There are two possibilities to explain this. One possibility is that the redshift of the Ly$\alpha$ emitters are out of our NB2288 narrowband window. The filter used by KC97 has a central wavelength of 4229\AA\ with a filter width of 70\AA. The Ly$\alpha$ emission that can enter into the window should be in the range $2.45<{\rm z}<2.50$, and the corresponding H$\alpha$ line will come to $2.261~\mu{\rm m} - 2.294~\mu{\rm m}$. As is shown in Fig.\ref{fig1}, almost half of the shorter wavelength range ($2.261~\mu{\rm m}-2.28~\mu{\rm m}$) is out of our survey window. However, if the Ly$\alpha$ emitters are all such objects, they would not be associated with, but slightly foreground to 4C~23.56.

Another explanation is a large error in the Ly$\alpha$ EW estimate, or spurious detections.  Due to the lack of $B$-band data which samples the continuum just redwards of Ly$\alpha$, they had to use a crude UV continuum estimate. If the EW of their Ly$\alpha$ emitters are much smaller than their quoted values, our non-detection of H$\alpha$ counterparts could be understood.  An independent Ly$\alpha$ emitter search is necessary to check the reliability of their selection of NB excess objects. We are in the progress of carrying out such a Ly$\alpha$ imaging survey for this field. Its results will be published elsewhere (Tanaka~I. et al., in preparation).

\subsection{Objects in \citet{kajisawa06b}}
\citet{kajisawa06b} took the $JHK$ images of the 4C~23.56 field and found a significant excess of ``{\it JHK}'' objects (see also, \cite{kodama07}). We checked their H$\alpha$ excess in our catalog. Though we could not identify their {\it b-JHK} objects, several Distant Red Galaxies (DRGs) were identified. The color excess of their DRGs is shown in Fig.\ref{fig12} (blue filled circles). Excluding 4C~23.56, two objects are identified in our HAE sample. Another two objects are identified as KC97's EROs. One of them shows emission-line excess (\#70 in KC97). Two other objects do not show detectable emission-line excess.

\subsection{An Object in \citet{stockton04}}
\citet{stockton04} reported an object in 4C~23.56 having very red color and a disk-like morphology. The SED fitting result indicates the strong preference for the single-burst model with fairly old age of $\sim2$Gyr at the redshift of the radio galaxy. The result suggests that the galaxy should be inactive \citet{stockton07}, and no H$\alpha$ emission is expected.

\begin{figure*}
\begin{center}
\FigureFile(120mm,100mm){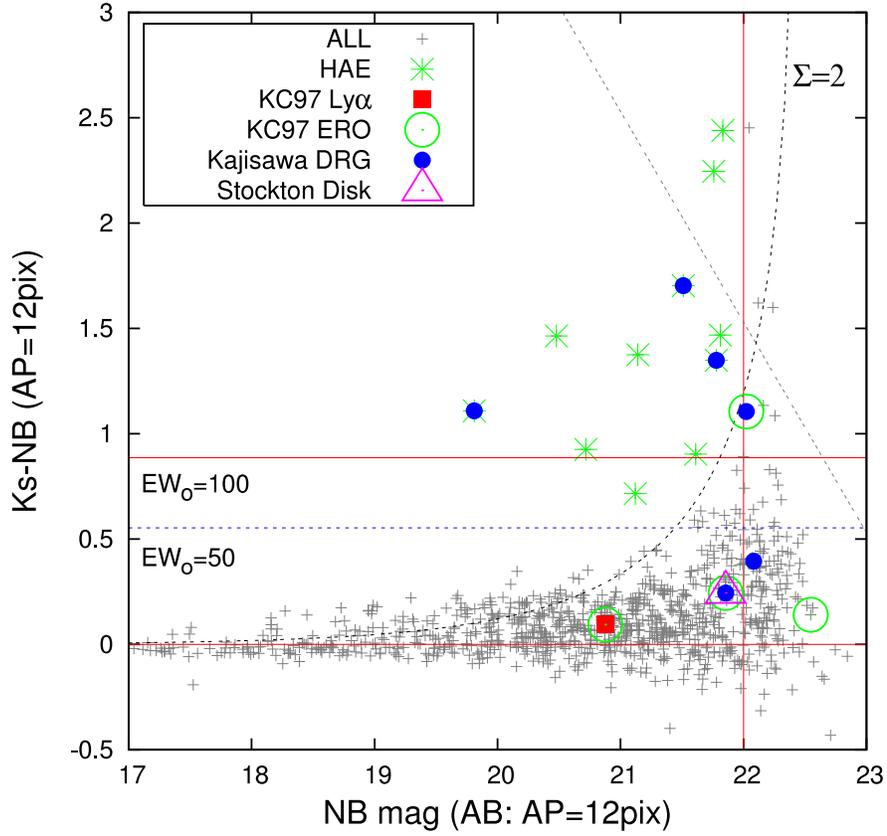}
\end{center}
\caption{Color excess of objects studied by others in the 4C~23.56 field. Only objects identified in our catalog are shown here. The grey plus marks are all objects in the field of view, and the asterisks are the HAEs. A red filled box and 4 green open circles each indicate a Ly$\alpha$ emitter and the EROs from KC97 that are in our catalog. Blue filled circles indicate the DRG sample from \citet{kajisawa06b}, and a magenta open triangle is the object studied by \citet{stockton04}. Ignoring the radio galaxy, three DRGs/EROs show the emission-line excess ($K_{\rm s}-{\rm NB}>1$), and they are also plotted in Fig.\ref{fig6}.\label{fig12}}

\end{figure*}

Their object is actually identified as an ERO in KC97 which is a DRG in \citet{kajisawa06b}. In Fig.\ref{fig12} we show the color excess of the object (magenta open triangle). As expected, the object does not show signs of excess emission from H$\alpha$.

%
%

\newpage
%
%
%
%

%
%

%
%

%
%

%
%

%
%
%
%

%
%

%
%

%
%

%
%

%
%

\clearpage
%
%
\setcounter{footnote}{0}
\begin{table*}
\begin{center}
\caption{Observational properties of candidate H$\alpha$ emitters.\label{tbl1}}
\begin{tabular}{ccccccccc}
\hline\hline
HAE ID & RA(J2000) & DEC(J2000) & \multicolumn{1}{c}{$K_{\rm s}$\footnotemark{}} & \multicolumn{1}{c}{EW$_{o}$\footnotemark{}} & \multicolumn{1}{c}{F(H$\alpha$)\footnotemark{}} & \multicolumn{1}{c}{F($24~\mu{\rm m}$)\footnotemark{}} & \multicolumn{1}{c}{redshift\footnotemark{}} \\
\hline
153 & 21:07:16.96 & 23:30:30.5  &  $21.82\pm0.07$  & 327 &  $3.64\pm0.77$ & $78.9$ & 2.4879 \\
354 & 21:07:21.03 & 23:31:13.9  &  $21.23\pm0.05$  & 105 &  $2.47\pm0.76$ & $134.9$ & 2.4865 \\
356 & 21:07:15.70 & 23:31:11.9  &  $22.78\pm0.12$  & 282 &  $1.35\pm0.50$ & $125.9$ & $\cdot\cdot\cdot$ \\
393\footnotemark{} & 21:07:21.48 & 23:31:19.5  &  $21.72\pm0.06$  &  51 &  $0.81\pm0.55$ & $42.4$ & $\cdot\cdot\cdot$ \\
431 & 21:07:15.93 & 23:31:27.5  &  $22.17\pm0.08$  & 221 &  $1.96\pm0.62$ & $156.6$ & $\cdot\cdot\cdot$ \\
479\footnotemark{} & 21:07:21.63 & 23:31:41.2  &  $22.37\pm0.08$  & 956 &  $4.25\pm0.72$ & $131.8$ & $\cdot\cdot\cdot$ \\
491\footnotemark{} & 21:07:14.80 & 23:31:45.1  &  $19.49\pm0.02$  & 122 & $13.95$ & $4630$ & 2.4852 \\
500\footnotemark{} & 21:07:21.84 & 23:31:41.8  &         $>23.5$    & $>780$ & $>0.8 $ & $\cdot\cdot\cdot$ & $\cdot\cdot\cdot$ \\
511\footnotemark{} & 21:07:22.58 & 23:31:43.4  &  $22.98\pm0.12$  & 300 &  $1.18\pm0.46$ & $156.5$ & $\cdot\cdot\cdot$ \\
526\footnotemark{} & 21:07:21.78 & 23:31:49.6  &  $20.76\pm0.04$  &  95 &  $3.45\pm0.94$ & $119.2$ & 2.4872 \\
543 & 21:07:17.61 & 23:31:50.5  &         $>23.5$    & $>505$ & $>0.8 $ & $\cdot\cdot\cdot$ & $\cdot\cdot\cdot$ \\
\hline
\end{tabular}
\end{center}
\footnotetext{}{1. the SExtractor MAG\_AUTO output in AB mag.}\\
\footnotetext{}{2. rest-frame equivalent width based on the ($K_{\rm s}-$NB2288) values by MAG\_AUTO output.}\\
\footnotetext{}{3. continuum subtracted narrowband flux in 10$^{-16}$~erg~sec$^{-1}$~cm$^{-2}$.}\\
\footnotetext{}{4. Spitzer MIPS $24~\mu{\rm m}$ flux in $\mu {\rm Jy}$ from the DAOPHOT photometry.}\\
\footnotetext{}{5. see \S3.1.}\\
\footnotetext{}{6. a possible low redshift interloper based on the $BJK$ criterion (\S3.1). }\\
\footnotetext{}{8. 4C~23.56. Its $K_{\rm s}$ magnitude is uncertain due to stray light (see \S2.1).}\\
\footnotetext{}{7,9,10,11. galaxies in the ``Eastern Clump'' (Fig.\ref{fig7}; see \S3.3).}\\
\end{table*}

\setcounter{footnote}{0}
%
%
\begin{table*}
\begin{center}
\caption{The estimated basic properties of the H$\alpha$ emitters.\label{tbl2}}
\begin{tabular}{ccccccccc}
\hline\hline
HAE ID & $\log({\rm M/M_{\solar}})$ & \multicolumn{1}{c}{$L_{\rm H\alpha}$\footnotemark{}} & SFR$_{\rm H\alpha}^{\rm raw}$ & \multicolumn{1}{c}{SFR$_{\rm H\alpha}^{\rm corr}$\footnotemark{}} & $\log(L_{24~\mu{\rm m}}/\L_{\solar})$ & \multicolumn{1}{c}{SFR($24~\mu{\rm m}$)\footnotemark{}}\\
\hline
153 & $11.124\pm0.235$ & $14.4\pm3.0$ &  67.13  & 187.96 & 12.533 & 410 \\
354 & $11.430\pm0.225$ & $9.75\pm3.02$ &  45.57  & 355.45 & 12.792 & 743 \\
356 & $10.635\pm0.260$ & $5.34\pm1.96$ &  24.94  & 122.21 & 12.758 & 688 \\
393 & $11.178\pm0.230$ & $3.19\pm2.16$ &  14.92  & 14.92   & 12.237 & 207  \\
431 & $10.948\pm0.243$ & $7.74\pm2.45$ &  36.17  & 177.23 & 12.864 & 877 \\
479 & $10.844\pm0.240$ & $16.75\pm2.82$ & 78.30  & 383.67  & 12.781 & 724 \\
491\footnotemark{} & $<12.320$\footnotemark{} & $55.0\pm6.8$ & 257.15 & 1429.75\footnotemark{} & N/A\footnotemark{} & N/A\footnotemark{} \\
500 & $<10.164$        & $>4.63$         & $>24.85$& $\cdot\cdot\cdot$ & $\cdot\cdot\cdot$ & $\cdot\cdot\cdot$ \\
511 & $10.531\pm0.262$ & $4.65\pm1.8$ & 21.72  & 106.43 & 13.091 & 876 \\
526 & $11.667\pm0.220$ & $13.6\pm3.7$ & 63.67  & 311.98 & 12.869 & 647 \\
543 & $<10.164$        & $>4.13$         & $>19.28$& $\cdot\cdot\cdot$ & $\cdot\cdot\cdot$ & $\cdot\cdot\cdot$ \\
\hline
\end{tabular}
\end{center}
\footnotetext{}{1. Extinction-uncorrected values in unit of $10^{42}$~erg~s$^{-1}$.}\\
\footnotetext{}{2. in ${\rm M}_{\solar}~{\rm yr}^{-1}$. See text for detail of the extinction correction.}\\
\footnotetext{}{3. {\it Spitzer} MIPS $24~\mu{\rm m}$-based SFR in ${\rm M}_{\solar}~{\rm yr}^{-1}$.}\\
\footnotetext{}{4,5. 4C~23.56. Its mass estimate is unreliable due to the uncertainty in its $K_{\rm s}$ magnitude (see \S2.1). If we use the value by KC97, the estimated stellar mass is 11.67.}\\
\footnotetext{}{6. Affected by an AGN contribution. The SFR value is quite uncertain.}\\
\footnotetext{}{7,8. Affected by an AGN contribution. The $L_{24~\mu{\rm m}}$ conversion from the observed MIPS flux is impossible due to the limitation of the Chary and Elbaz (2001) model.}\\
\end{table*}

\end{document}